%% file: ms_arxiv.tex
\documentclass[apj]{emulateapj}
\usepackage{graphics}
\usepackage{enumitem}
\usepackage{amsmath}
\usepackage{xcolor}
\usepackage{soul}

\usepackage{lineno}

\def\simlt{\lower.5ex\hbox{$\; \buildrel < \over \sim \;$}}
\def\simgt{\lower.5ex\hbox{$\; \buildrel > \over \sim \;$}}

\def\msun{{$\rm M_\odot$}}

\begin{document}

\shortauthors{TEMIM ET AL.}

\shorttitle{Progenitor of G292}

\title{SNR G292.0+1.8: A Remnant of a Low-Mass Progenitor Stripped-Envelope Supernova}

 \author{Tea Temim\altaffilmark{1}, Patrick Slane\altaffilmark{2}, John C. Raymond\altaffilmark{2}, Daniel Patnaude\altaffilmark{2}, Emily Murray\altaffilmark{1}, Parviz Ghavamian\altaffilmark{3}, Mathieu Renzo\altaffilmark{4}, Taylor Jacovich \altaffilmark{2,5}}

\altaffiltext{1}{Princeton University, 4 Ivy Ln, Princeton, NJ 08544,  temim@astro.princeton.edu}
\altaffiltext{2}{Harvard-Smithsonian Center for Astrophysics, 60 Garden Street, Cambridge, MA 02138, USA}
\altaffiltext{3}{Towson University, Department of Physics, Astronomy and Geosciences, Towson, MD 21252, USA}
\altaffiltext{4}{Center for Computational Astrophysics, 162 5th Ave, New York, NY 10010, USA}
\altaffiltext{5}{The George Washington University, Department of Physics
725 21st Street, NW Washington, DC 20052}

\begin{abstract}

We present a study of the Galactic supernova remnant (SNR) G292.0+1.8, a classic example of a core-collapse SNR that contains oxygen-rich ejecta, circumstellar material, a rapidly moving pulsar, and a pulsar wind nebula (PWN). We use hydrodynamic simulations of the remnant evolution to show that the SNR reverse shock is interacting with the PWN and has most likely shocked the majority of supernova ejecta. In our models, such a scenario requires a total ejecta mass of $\lesssim \rm 3\: M_{\odot}$ and implies that there is no significant quantity of cold ejecta in the interior of the reverse shock. In light of these results, we compare the estimated elemental masses and abundance ratios in the reverse-shocked ejecta to nucleosynthesis models and find that they are consistent with a progenitor star with an initial mass of 12--16~$\: \rm M_{\odot}$. We conclude that the progenitor of G292.0+1.8 was likely a relatively low mass star that experienced significant mass loss through a binary interaction and would have produced a stripped-envelope supernova explosion. We also argue that the region known as the ``spur" in G292.0+1.8 arises as a result of the pulsar's motion through the supernova ejecta and that its dynamical properties may suggest a line-of-sight component to the pulsar's velocity, leading to a total space velocity of $\sim600\: {\rm km\:s^{-1}}$ and implying a significant natal kick. Finally, we discuss binary mass loss scenarios relevant to G292.0+1.8 and their implications for the binary companion properties and future searches.

\end{abstract}

\section{Introduction} \label{intro}

Core-collapse (CC) supernovae (SNe), the cataclysmic explosions of massive stars ($>$8~$\rm \:M_\odot$), are responsible for the production and distribution of heavy elements that enrich their surroundings and profoundly affect the interstellar media in which they evolve. They leave behind compact objects whose high densities and magnetic field strengths represent matter under the most extreme conditions known. While significant progress has been made in the theoretical understanding of how massive stars explode, many questions remain about the precise mechanism and how the explosions will vary with progenitor star properties and the mass loss experienced prior to collapse \citep[see review by][]{burrows21}.

Indeed, the SN types we observe are primarily related to the initial stellar mass and the nature of mass loss through stellar winds, eruptions, and mass transfer to a binary companion. While the progenitors of Type IIP SNe still have considerable hydrogen (H) envelopes, indicating little mass loss, and those of Type IIL/b SNe have some of their H envelopes intact, the Ib/Ic SNe are thought to originate from stripped-envelope progenitors that experienced significant mass loss either through strong stellar winds or binary interactions \citep[see review by][]{smith14}. The fraction of stripped-envelope SNe that originate from single-star winds versus lower mass stars in binary systems is still under debate, but evidence is mounting that the latter may dominate \citep[e.g.][]{prentice19}.

Observational signatures of young supernova remnants (SNRs) are directly shaped by the SN explosion properties, the composition and mass of the ejected material, and the nature of the circumstellar material (CSM) expelled in the late stages of the progenitor's life \citep{patnaude17}. SNRs therefore provide the means of resolving and directly probing the CSM structure and SN ejecta asymmetries and abundances. However, linking SNRs to progenitor properties and SN explosion sub-types is crucial for comparing their characteristics to theoretical expectations, but has proven to be a difficult challenge. Progenitor mass estimates for core-collapse SNRs have been made using observed abundance ratios \citep[e.g.][]{katsuda18} and surveys of surrounding stellar populations \citep[e.g.][]{diaz-rodriguez18,williams19,auchettl19}, but additional information about explosion properties has been obtained for only a few very young core-collapse SNRs \citep[e.g.][]{smith13,yang15,borkowski16,temim19}. A firm identification of the explosion sub-type for a Galactic core-collapse SNR has been made only for Cassiopeia~A through light echo observations \citep{krause08a}.

SNR G292.0+1.8 is one of the best studied Galactic SNRs that exhibits all of the characteristics that might be expected from a core-collapse explosion. It contains bright filamentary emission from oxygen-rich ejecta \citep{park07}, surrounding CSM emission and a circumstellar ring \citep{park02}, a pulsar that is rapidly moving away from the explosion center \citep{hughes01}, and a pulsar wind nebula (PWN) that is visible at X-ray and radio wavelengths \citep{hughes01,gaensler03}. In this work, we present dynamical simulations of the evolution of G292.0+1.8 and show that the inclusion of the central PWN and the pulsar's motion through the SNR can provide important constraints on the progenitor star and SN explosion. The paper is organized as follows: Section~\ref{properties} summarizes the observational properties of G292.0+1.8. The modeling of the SNR and PWN evolution is presented in Section~\ref{models}, comparison of the X-ray abundances with nucelosynthesis models in Section~\ref{abundmodel}, interpretation of the infrared and optical properties in Section~\ref{spur_nature}, and a summary and discussion of the results in Section~\ref{summary}. Finally, the main conclusions are summarized in Section~\ref{conclusion}.

\section{Observational Properties} \label{properties}

\begin{figure}
\center
\includegraphics[width=0.48\textwidth]{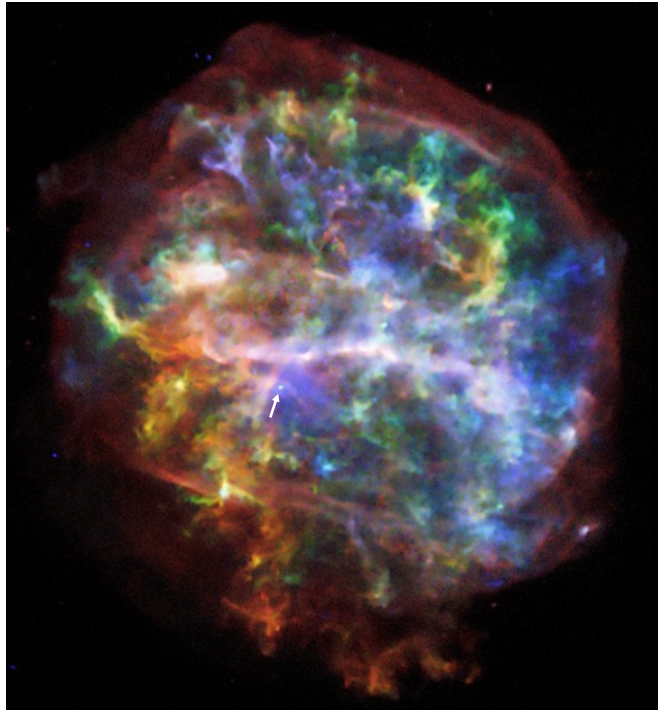}
\caption{\label{xray} \textit{Chandra} X-ray image of G292.0+1.8, showing the O and Ne emission in red (0.580-710 and 0.880-950~keV), Ne in orange (0.980-1.100~keV), Mg in green (1.280-1.430~keV), and Si and S in blue (1.810-2.050 and 2.400-2.620~keV). The PWN is seen in blue, displaced to the southeast of the shell center. The pulsar is indicated by the white arrow. The SNR radius is $\sim$265~\arcsec, corresponding to 7.7~pc for a distance of 6.0~kpc. Image credit: NASA/CXC/\citet{park07}.}
\end{figure}

\begin{figure*}
\center
\includegraphics[width=0.99\textwidth]{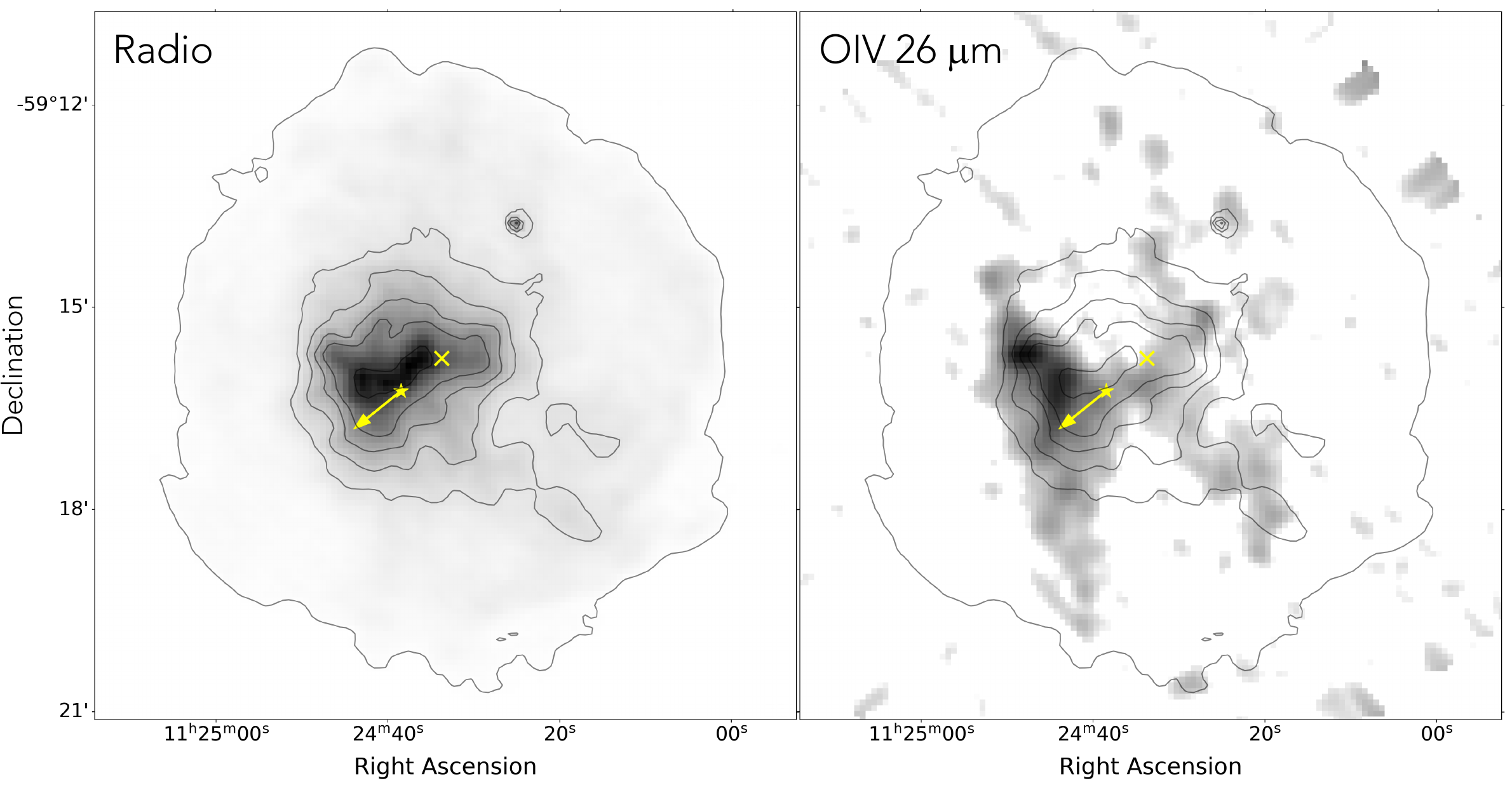}
\caption{\label{radio_ir}Left: A 6-cm radio image from the Australia Telescope Compact Array \citep{gaensler03}, overlaid with the contours form the outer SNR shell and the PWN. Right: A narrow band image of the \ion{O}{4}~25.89~\micron\ line from \textit{Spitzer} IRS \citep{ghavamian12}, overlaid with the radio contours shown on the left. The yellow \textit{X} marks the location of the explosion center \citep{winkler09}, and the star indicates the current location of the pulsar with the arrow pointing in the direction of the pulsar's motion. The outer contours of the SNR shell and PWN are approximately 265\arcsec\ and 130\arcsec, respectively. The pulsar is displaced from the center of the PWN towards the SE (the direction of the pulsar's motion). The distance from the pulsar to the outer edge of the PWN is $\sim$90\arcsec\ in the direction of the pulsar's motion and $\sim$160\arcsec\ in the opposite direction towards the NW.}
\end{figure*}

An image of SNR G292.0+1.8, as seen at X-ray wavelengths, is shown in Figure~\ref{xray} \citep{park07}. The remnant shell is $\sim$265\arcsec\ in size, corresponding to 7.7~pc at a distance of 6~kpc, and it exhibits a patchwork of bright knots and filaments of metal-rich ejecta. The adopted distance of 6~kpc is based on the lower limit of 6.2$\pm$0.9~kpc from the analysis of \ion{H}{1} emission and absorption towards the SNR \citep{gaensler03}, 5.4~kpc from the reddening of optical filaments \citep{goss79}, and 6.4$\pm$1.3~kpc from the dispersion measure of the pulsar \citep{camilo02}.
The CSM is distributed along an equatorial band, as well as in a more diffuse shell along the circumference of the SNR. The pulsar and its X-ray PWN are displaced to the southeast of the geometric center of the shell (see white arrow in Figure~\ref{xray}).  The radio image \citep{gaensler03} is shown in the left panel of Figure~\ref{radio_ir}, with the contours outlining the outer boundary of the SNR shell and the PWN overlaid. Optical observations of
G292.0+1.8 revealed a network of fast (up to $\sim$3500~$\rm km\:s^{-1}$) ejecta filaments distributed in bi-polar lobes oriented north-south and extending as far as the outer boundary of the SN ejecta observed in X-rays \citep{murdin79,winkler06,ghavamian05, winkler09}. Proper motion measurements of these optical filaments were used to estimate the center of explosion (see Figure~\ref{radio_ir}) and a kinematic age of $\sim$3000~years for the remnant, assuming no deceleration of the filaments \citep{ghavamian05,winkler09}. A region of bright optical filaments with lower velocities ($\leq1800~\rm km\:s^{-1}$) known as the ``spur" is located in the southeastern quadrant of the SNR \citep{goss79, murdin79, ghavamian05, winkler06}. The infrared (IR) continuum emission from dust generally traces the CSM distribution seen in X-rays \citep{lee09,ghavamian12}, but the IR line emission from oxygen and neon in particular (see right panel of Figure~\ref{radio_ir}) is coincident with the spur \citep{lee09,ghavamian12}.

\subsection{Ejecta Abundances and Total Mass} \label{ejecta}

The abundances of the supernova ejecta in G292.0+1.8 have been derived from X-ray observations and shown to be dominated by O, Ne, and Mg, with much smaller quantities of Si, S, and Fe \citep{gonzalez03,park04,kamitsukasa14,bhalerao19}. Comparisons of the abundances to core-collapse nucleosynthesis models have generally led to progenitor mass estimates of 25--40~$M_{\odot}$ \citep{hughes94,gonzalez03,yang14,kamitsukasa14}. A more recent study by \citet{bhalerao19} used deep \textit{Chandra} observations to conduct a survey of elemental abundances across the SNR and derive total integrated masses of the observed elements (summarized in their Table~4). Their comparison of abundances for each element to core-collapse nucleosynthesis models of \citet{nomoto06} led to different progenitor mass constraints, so they provide a wider progenitor mass range of 13--30~$\rm M_{\odot}$.

An anomaly noted by these previous X-ray studies is the very weak
emission from Fe that leads to comparatively low total Fe mass in the
SNR (0.03$\pm$0.01~$\rm M_{\odot}$ estimated by \citealt{bhalerao19}),
lower than would be expected from any core-collapse nucleosynthesis
models. This led numerous authors to suggest that the reverse shock
has not yet penetrated deep enough into the interior of the SNR, leaving the
majority of the Fe still unshocked \citep{park04,bhalerao19}. 
 Previous
studies have also noted an anomalously large overabundance of Ne with
respect to oxygen that cannot be reconciled with the current models.
The overabundance was observed both in X-ray \citep[e.g][]{park04} and
IR observations \citep{ghavamian09}. \citet{bhalerao19} find the total
combined mass of O, Ne, Mg, Si, S, and Fe emitting in the X-rays to be
1.3$^{+0.5}_{-0.3}$~$\rm M_{\odot}$, but they argue that as much as
4~$\rm M_{\odot}$ could still be unshocked by the reverse shock,
including the majority of Si, S, and Fe whose abundances appear to be
low in comparison to nucleosynthesis models of high mass stars.

\subsection{Circumstellar Material}

Emission from CSM material has been identified in G292.0+1.8 based on solar abundances found in the spectral fits to X-ray observations \citep[e.g.][]{park02,gonzalez03}. The morphology of the CSM resembles thin filaments of soft X-ray emission and a more diffuse soft X-ray shell around the perimeter of the SNR. This emission is seen in red in Figure~\ref{xray} \citep{park07}. A distinct ring of CSM material is found in the equatorial region of the SNR and has been detected at X-ray \citep{park07}, optical \citep{ghavamian05}, and IR wavelengths \citep{lee09,ghavamian12}. Through their detailed X-ray spectral analysis of the deep \textit{Chandra} observations, \citet{bhalerao19} estimate the total mass of shocked CSM material to be $13.5^{+1.7}_{-1.4}f^{1/2}d_6^{3/2}\rm \: M_{\odot}$, where $f$ is the filling factor and $d$ is the distance, assumed here to be 6~kpc.
Out of the total estimated CSM mass, $1.7\pm0.1\rm\:M_{\odot}$ is
attributed to the equatorial ring. \citet{lee10} used the same X-ray
data set to measure the radial variations in the properties of the CSM
component and found a decreasing density profile with a wind density of $n_H=0.1-0.3\:\rm cm^{-3}$ at the current outer radius of the SNR, that they concluded is indicative of a red supergiant (RSG) wind profile. Based on the measured SNR radius and assuming expansion into an RSG wind profile (with a mass loss rate of $3\times10^{-5}\:\rm M_{\odot}\:yr^{-1}$ and a wind velocity of 15~$\rm km\:s^{-1}$), \citet{chevalier05} calculated the total swept-up CSM mass for G292.0+1.8 to be $\sim$14~$\rm M_{\odot}$, in agreement with the X-ray-measured swept-up mass of \citet{bhalerao19}.

\subsection{Pulsar and the PWN} \label{pwn_obs}

SNR G292.0+1.8 hosts an energetic pulsar PSR J1124--5916, with a period of 135~ms, a spin-down luminosity of $1.2\times37\rm \: erg\:s^{-1}$, and a characteristic age of 2900~years \citep{camilo02}. The pulsar is displaced by approximately 46\arcsec\ (1.34~pc for a distance of 6~kpc) to the southeast of the optically-determined center of expansion, assumed to be the center of explosion. This displacement implies that the pulsar is moving to the southeast (see yellow arrow in Figure~\ref{radio_ir}) with a transverse velocity of $\sim500\rm\:km\:s^{-1}$. 
An X-ray-emitting PWN with an approximate radius of $\sim$~65\arcsec\ surrounds the pulsar \citep{hughes01}. As seen in Figure~\ref{radio_ir}, the PWN is more extended at radio wavelengths with a radius of $\sim$~130\arcsec\ or 3.8~pc at 6~kpc. Interestingly, the pulsar is also displaced from the center of the radio PWN. The distance from the pulsar to the outer edge of the PWN is $\sim$~90\arcsec\ in the direction of the pulsar's motion and $\sim$~160\arcsec\ in the opposite direction towards the NW.

\subsection{Evolutionary Stage}

In their radio study of G292.0+1.8, \citet{gaensler03} concluded that the SNR reverse shock has already reached the PWN and began interacting with it. The main evidence for this was the absence of any obvious gap between the emission from the PWN and the emission from the SNR shell that they argued arose from reverse-shocked ejecta. \citep{bhalerao15} used the \textit{Chandra} High Energy Transmission Grating Spectrometer data to study the kinematics and spatial distribution of the reverse-shocked ejecta knots in G292.0+1.8 and concluded that the radius of the reverse shock is approximately 4~pc from the center of the SNR, very close to the edge of the PWN as seen at radio wavelengths. Their results also suggested that it is possible that the reverse shock and the PWN have already collided. However, as described above, the abundance anomalies related to the low Fe content in the SNR led other authors to conclude that the reverse shock has not reached the majority of the ejecta, in which case it would not already be interacting with the PWN.

\section{Modeling the SNR \& PWN Evolution} \label{models}

\subsection{1-D Semi-Analytical Model} \label{1d}

In order to constrain the evolutionary stage of G292.0+1.8 and the
mass of its progenitor star, we used 1-D semi-analytical and 2-D
hydrodynamic (HD) models for the evolution of a PWN inside an SNR to
reproduce the observational properties described in
Section~\ref{properties}. The 1-D semi-analytical model is described
in detail in \citet{gelfand09}. It simultaneously calculates the
dynamical evolution of the SNR and PWN system, as well as the
evolution of the PWN's broadband spectrum and magnetic field as a
function of the SNR age. The input parameters to the model include the
SN ejecta mass $M_{ej}$, the ambient density $n_0$, SN explosion
energy $E_{51}$, the pulsar breaking index $n$, and the pulsar's
spin-down timescale $\tau_0$. These parameters were varied to
reproduce the observationally-constrained values for the pulsar's
spin-down luminosity and
characteristic age, the broadband spectrum of the PWN, as well as the observed radii of the SNR and PWN. Given these constraints, the model also predicts the SNR age and the magnetic field of the PWN. The SN ejecta is assumed to have an outer envelope whose density drops off as $\rho\propto r^{-12}$ \citep{truelove99}. We note that the model employs a constant density ambient medium.

\input{tab1.tex}


As we will show in the next section, our subsequent HD model simulations show that the observed displacements of the pulsar's position from the center of the radio PWN can only be achieved if the SNR reverse shock has begun interacting with the PWN in the southeast. We therefore further constrained the 1-D model by requiring the SNR reverse shock radius to be equal to the PWN radius along the axis aligned with the pulsar's direction of motion (equal to 3.8~pc). We then produced a set of 1-D models that can reproduce the current size of the SNR, reverse shock, and the PWN, as well as the PWN's radiative properties. We used input ejecta masses ranging from 1.4~$\rm \: M_{\odot}$ to 5.0~$\rm \: M_{\odot}$, as masses higher than this could not reproduce the PWN properties discussed in Section~\ref{hydro}. We then adjusted the ambient density and explosion energy accordingly, such that the ambient density is at the lowest value that would still reproduce the observed parameters for the SNR. This approach was used in order to keep the total swept-up CSM mass as low as possible for each run, since the total estimated CSM mass from X-ray observations is only $13.5^{+1.7}_{-1.4}f^{1/2}d_6^{3/2}\rm \: M_{\odot}$ \citep{bhalerao19}.

\input{tab2.tex}


\begin{figure}
\center
\includegraphics[width=0.47\textwidth]{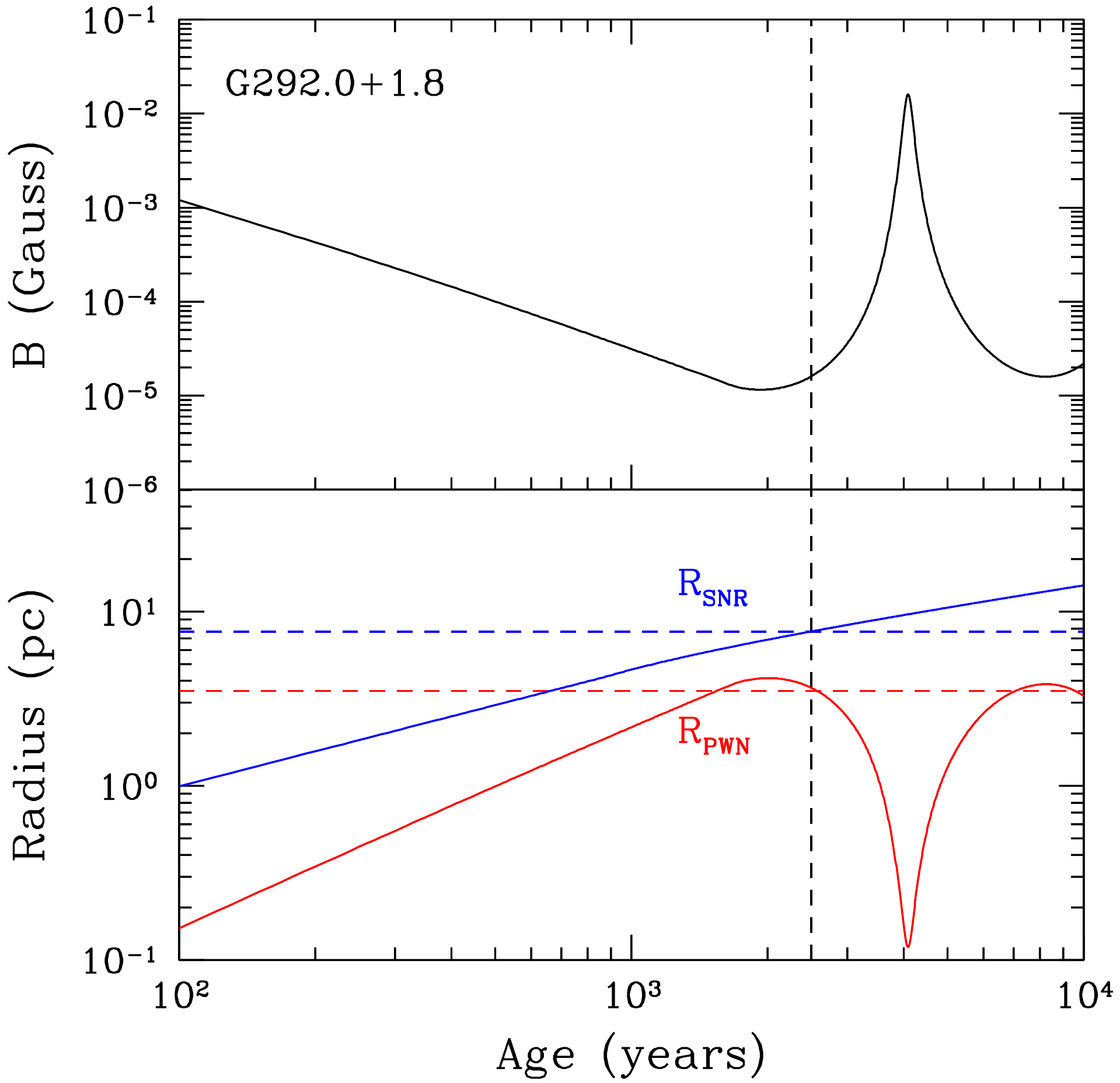}
\includegraphics[width=0.47\textwidth]{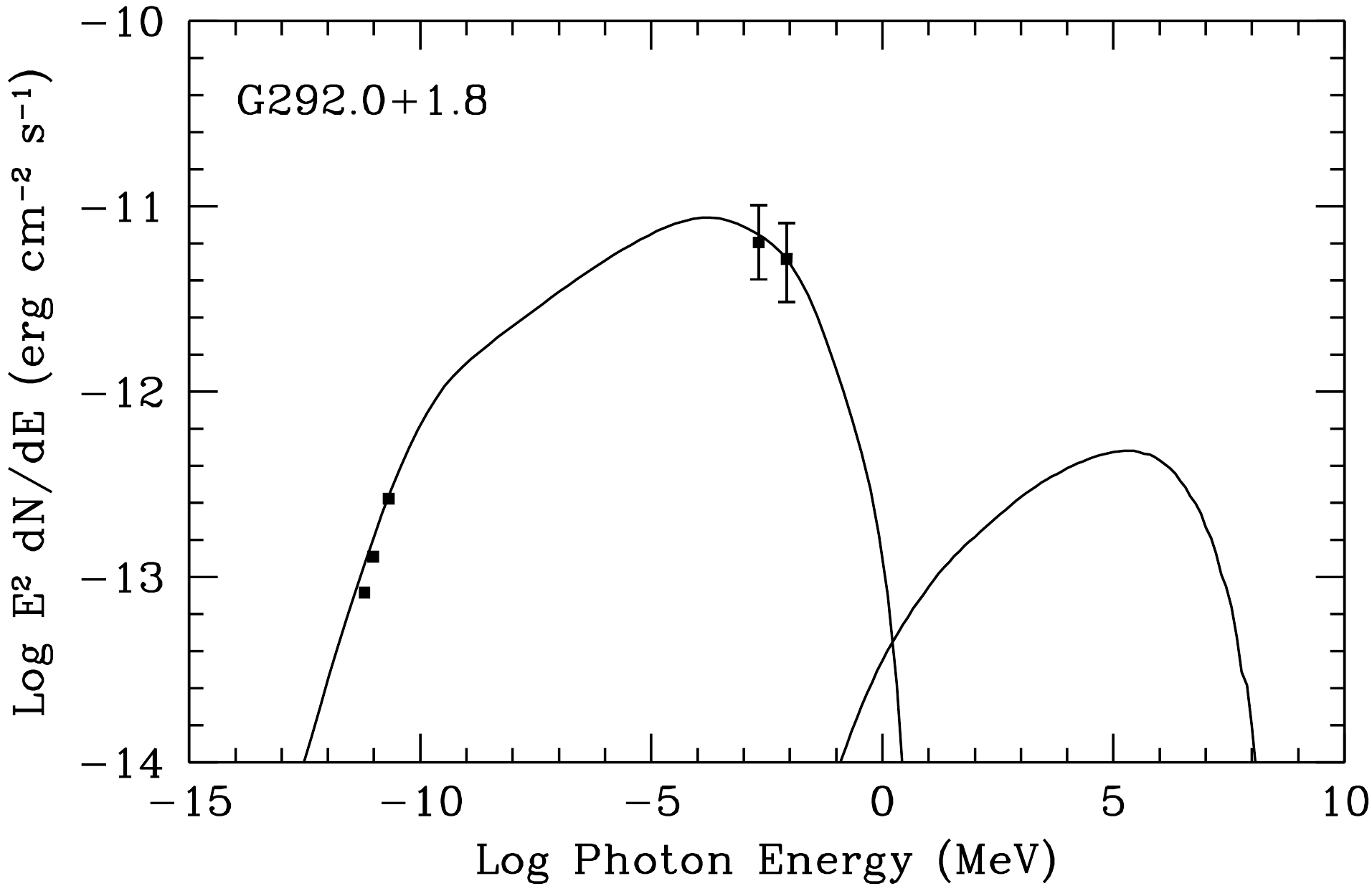}
\caption{\label{1dmodel}
Top: Time evolution of the magnetic field (upper panel) and the SNR and PWN radii (lower panel) for G292.0+1.8 produced by the 1D model described in Section~\ref{1d}. The input parameters for the model are those for Run~2 listed in Table~\ref{model_param}. The vertical dashed line represents an age of 2500~years. The observed radius of the SNR and the PWN are indicated by the horizontal blue and red dashed lines, respectively.
Bottom: 1D-model broadband spectrum of the PWN in G292.0+1.8 at the age of 2500~years corresponding to the model parameters listed in Table~\ref{model_param} with the radio \citep{gaensler03} and X-ray \citep{hughes01} fluxes from the PWN overlaid.}
\end{figure}

The set of models that successfully reproduce the observed properties is listed in Table~\ref{1Dparam}. The resulting age for all the runs is around 2500~years. This is lower than the derived kinematic age of $\sim$~3000~years \citep{ghavamian05,winkler09}, but a lower age is not unexpected since the filaments have likely experienced some deceleration in the SNR lifetime. The plots showing the time evolution of the magnetic field and the SNR and PWN radii for Run~3, using the parameters listed in Table~\ref{model_param}, are shown in Figure~\ref{1dmodel}. The broadband spectrum at the final age of 2500~years is shown in the bottom panel of Figure~\ref{1dmodel}. Again, we note that all runs in Table~\ref{1Dparam} produce similar plots that match the observational constraints. For all models, the reverse shock has already begun compressing the PWN, reducing its radius to the present size.

We also examine the total swept-up CSM mass in the models which we calculated by multiplying the input ambient density by the volume of the SNR. Since G292.0+1.8 is thought to still be interacting with CSM material \citep{lee10}, we assumed that all of the swept-up material can be attributed to the CSM.
The resulting CSM masses for all runs are listed in Table~\ref{1Dparam}. The table also includes the ZAMS mass for each run, calculated by adding the total CSM mass, the mass of the ejecta, and $1.6\:\rm \: M_{\odot}$ that we assume is contained in the neutron star.
As we increase the ejecta mass in the model, the required ambient density and swept up CSM mass also increase, primarily due to the constraint that the reverse shock needs to be at the PWN boundary, equivalent to a forward shock to reverse shock radius ratio of about two. We can see from Table~\ref{1Dparam} that runs 1--2 are  roughly consistent with the observed CSM mass, while runs 3--5 exceed the observed values. We also note that ejecta masses in runs 1--4 are too low to be consistent with single star progenitors, and instead would be more consistent with lower mass binary-stripped progenitors. For runs 3 and 4, the ZAMS mass is implausibly high for such a low ejecta mass, The ejecta and CSM masses for Run 5 may be plausible for a very high mass ($>$40~$\rm \: M_{\odot}$) single star stripped-envelope progenitor \citep[e.g., see][]{sukhbold16}, but the associated CSM mass greatly exceeds the mass of the swept-up CSM measured from observations.

\subsection{2-D Hydrodynamic Model} \label{hydro}

\begin{figure*}
\center
\includegraphics[width=\textwidth]{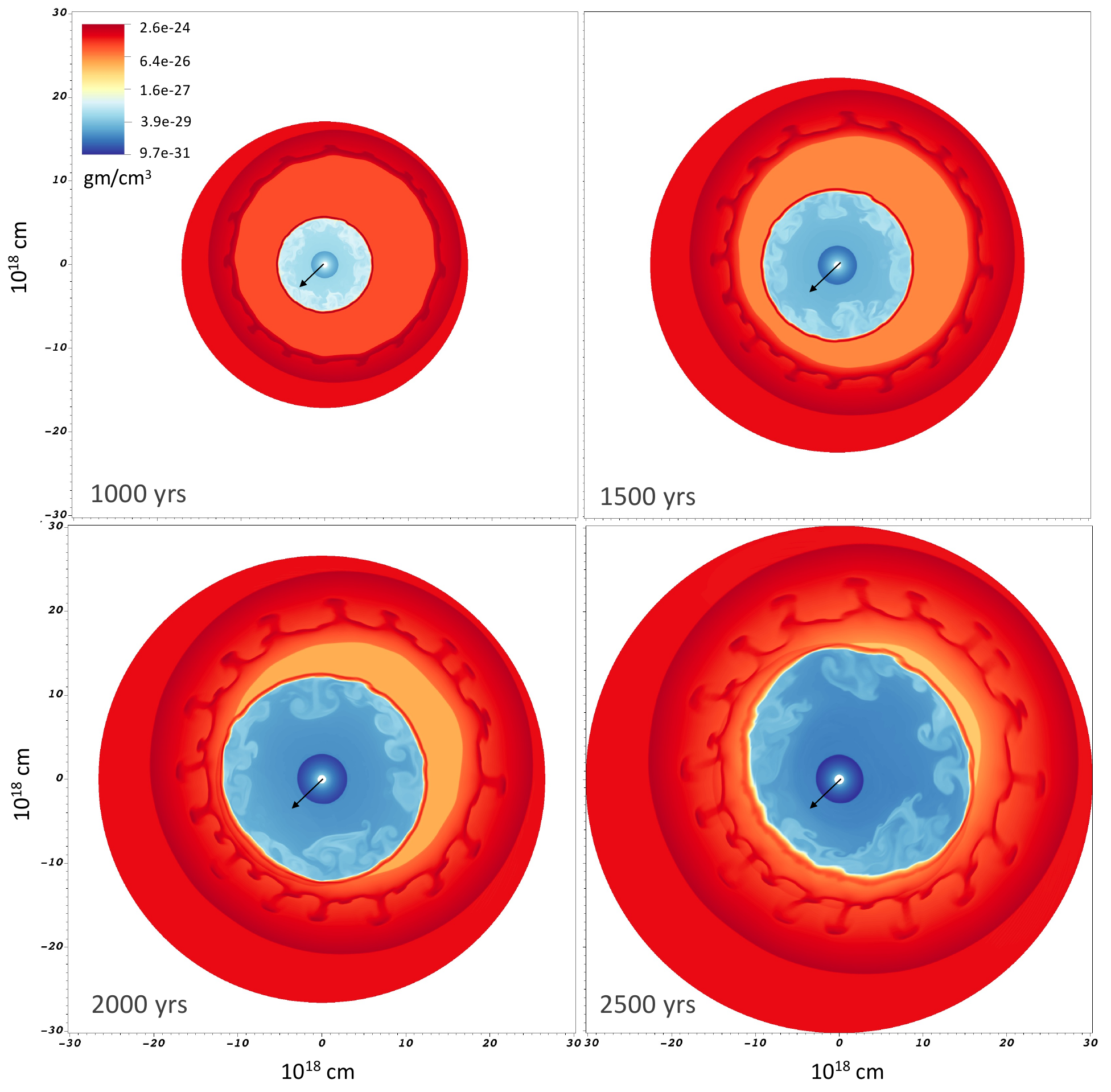}
\caption{\label{hydro_grid}Densities resulting from the HD simulations of G292.0+1.8 at 1000, 1500, 2000, and 2500~years. The input parameters for the simulation correspond to Run~3 in Tables~\ref{1Dparam} and \ref{model_param}. The pulsar is moving in the direction indicated by the black arrows, so the reverse shock initially impacts the PWN from this direction after 1500~years. By 2500~years, the reverse shock has interacted with the entire surface of the PWN. The pulsar remains in the center of the PWN until the reverse shock interaction begins, at which point it becomes displaced towards the SE (the direction of its motion), as observed in the radio image in Figure~\ref{radio_ir}.}
\end{figure*}

We used the final parameters from select 1-D models described in the previous section as inputs to 2-D HD models that account for the proper motion of the pulsar through the SNR. The HD model uses the VH-1 hydrodynamics code \citep{blondin01} that has been modified to simulate a moving pulsar and its PWN evolving inside an SNR \citep{temim15, kolb17}. The pulsar generates a PWN that is simulated as a bubble of relativistic gas powered by the pulsar's time-dependent spin-down luminosity \citep[for details see][]{kolb17}. The SNR is modeled as a self-similar wave expanding into the same ejecta density profile as used in the 1-D models (a power-law drop-off with index of 12), and a uniform ambient density. The simulations are carried out in the rest frame of the pulsar, but the SNR and ambient medium are given a velocity of $500\:\rm km\:s^{-1}$ to mimic the pulsar's observed velocity through the SNR.

\begin{figure}
\center
\includegraphics[width=0.42\textwidth]{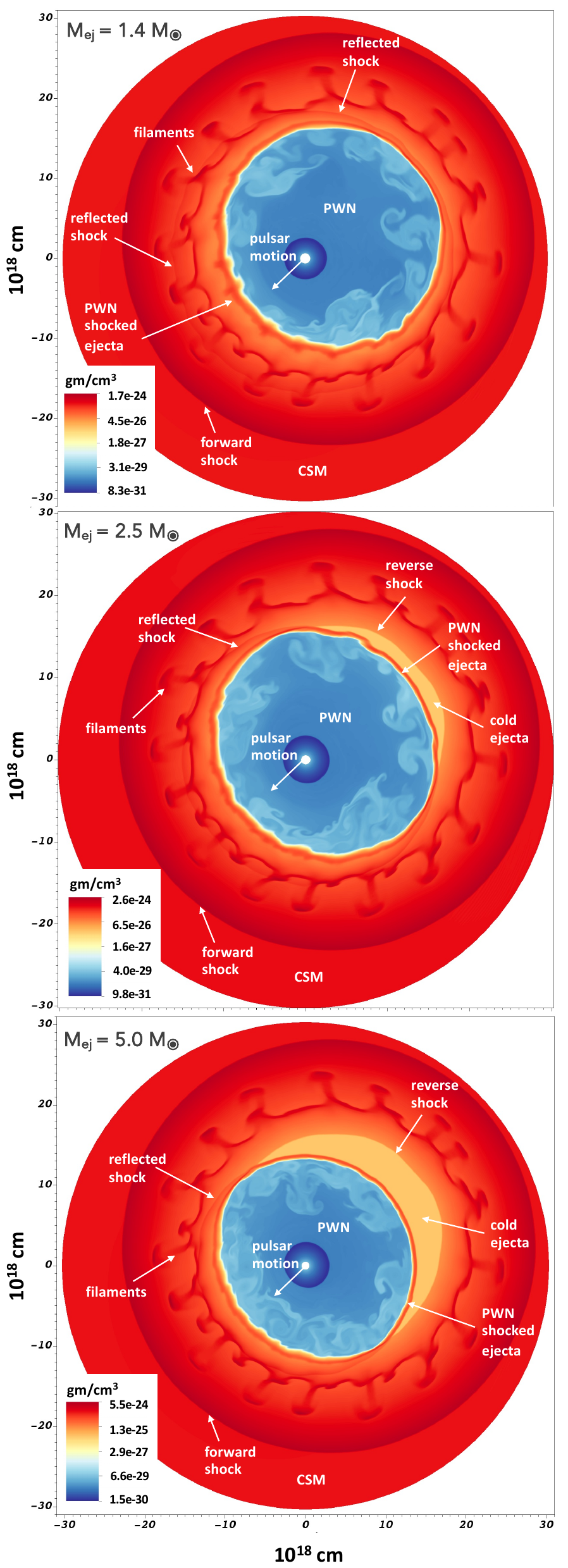}
\caption{\label{hydro_2panel}Density resulting from the HD simulations of G292.0+1.8 at an age of 2500~years. The input parameters for the simulations are listed in Tables~\ref{1Dparam} and \ref{model_param}, and correspond to Run~1 (top), Run~3 (middle) and Run~5 (bottom). The pulsar is moving towards the southeast. Various features discussed in Section~\ref{hydro} are labeled with white arrows. While the reverse shock interacted with the entire surface of the PWN for Run~1, it has not yet reached the northwestern side of the PWN for the higher-ejecta-mass models.}
\end{figure}

The simulation corresponding to Run~3 in Tables~\ref{1Dparam} and \ref{model_param} is shown in Figure~\ref{hydro_grid}. The middle panel of Figure~\ref{hydro_2panel} shows the final panel of Figure~\ref{hydro_grid} labeled to indicate the various morphological structures. The four panels show the density evolution at 1000, 1500, 2000, and 2500 years, with the direction of the pulsar's motion indicated by the black arrows. At 1000~years, it can be seen that the pulsar and the PWN are displaced from the center of the SNR shell. Because the initial expansion speed of the PWN is much higher than that of the pulsar, the pulsar remains located near the center of the PWN.

At 1500~years, the reverse shock has almost reached the PWN in the southeast, since the PWN has moved closer to the shock in this direction due to the pulsar's motion. At this age, the formation of large-scale Rayleigh-Taylor filaments as the reverse shock propagates through the ejecta is also evident. The thin red shell surrounding the blue-colored PWN is the swept-up, inner SN ejecta that is shocked by the PWN. At 2000~years, the reverse shock has encountered the southeastern hemisphere of the PWN and as they collide, a reflected shock (labeled in Figure~\ref{hydro_2panel}) is seen to form and propagate in the direction away from the explosion center and through the large-scale filaments of ejecta in the southeast. By the final age of 2500~years, the majority of the PWN's surface has been reached by the reverse shock, which also means that most of the SN ejecta have been shocked at this stage.

\begin{figure*}
\center
\includegraphics[width=0.8\textwidth]{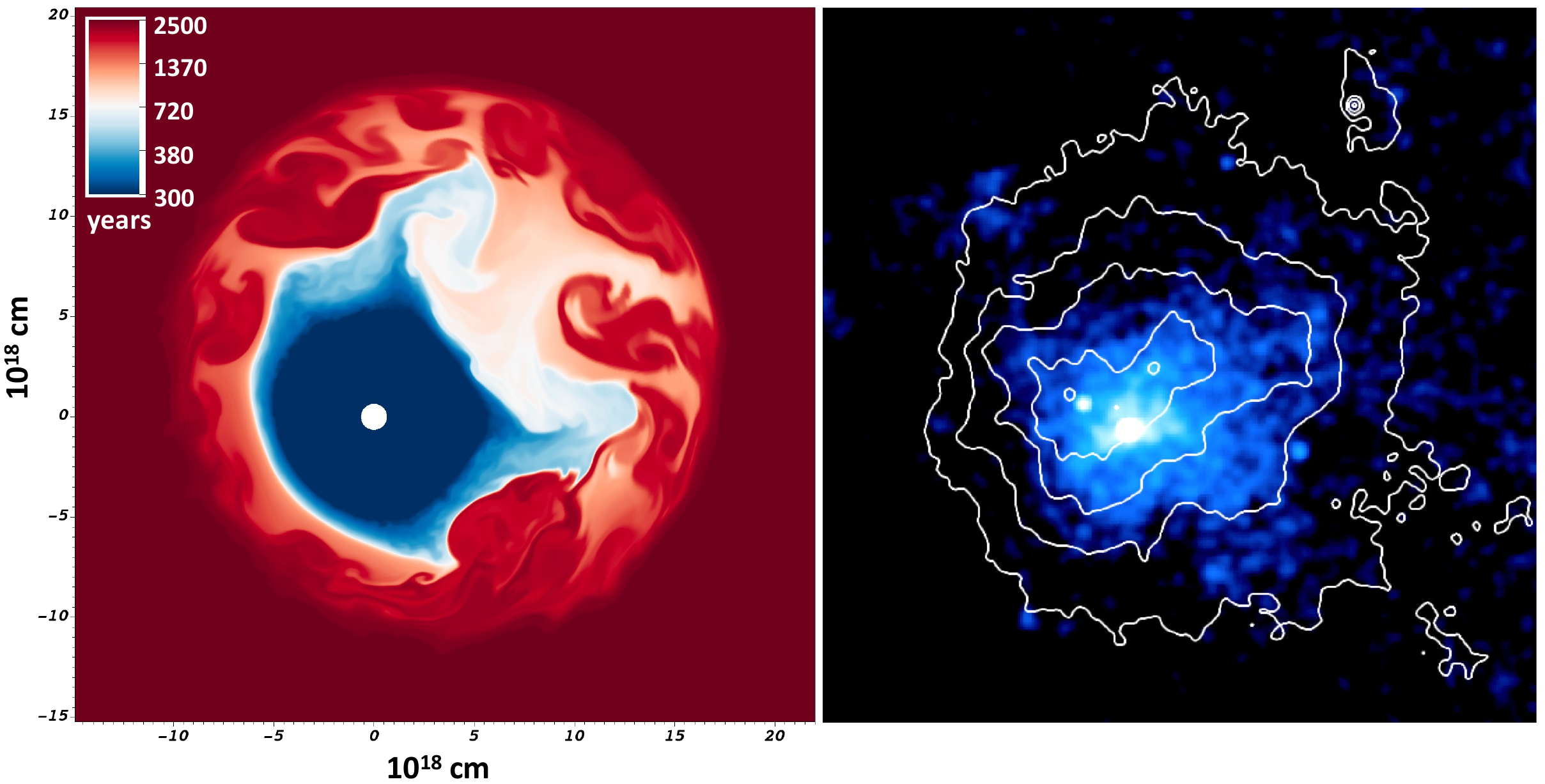}
\caption{\label{synch_life}Left: Map of the particle age within the PWN for the HD simulation using input parameters for Run~1 listed in Table~\ref{model_param} at an age of 2500~yrs. For a magnetic field of 13~$\mu G$ (Table~\ref{model_param}), the synchrotron lifetime of 3~KeV-emitting particles is $\sim$~450~yrs. Right: \textit{Chandra} X-ray in the 3-6~keV band showing the extent of the X-ray PWN, with the radio contours overlaid. The significantly smaller extent of the X-ray PWN with respect to the radio can be explained by the differences in particle ages across the PWN.}
\end{figure*}

An important thing to note is that only once the reverse shock begins compressing the PWN from the direction in which the pulsar is moving does the position of the pulsar become displaced from the center of the PWN towards the southeast. As the reverse shock continues to propagate inward towards the explosion center, the pulsar's displacement from the center of the PWN becomes larger.
The fourth panel of Figure~\ref{hydro_grid} clearly shows the displacement of the pulsar to the southeast of the PWN center at the final age of 2500~years.
This is consistent with the pulsar's observed position within the radio PWN, as described in Section~\ref{pwn_obs}.

In the top and bottom panels of Figure~\ref{hydro_2panel}, we also show the final density profile at 2500~years for an HD simulation with input parameters corresponding to runs~1 and 5 in Tables~\ref{1Dparam} and \ref{model_param}. For the model with the lowest ejecta mass (Run~1) of 1.4$\rm \: M_{\odot}$, the reverse shock has interacted with the entire surface of the PWN at 2500~years and all of the ejecta have been reverse-shocked. As the input ejecta mass increases, the radius of the PWN decreases at the final age since the PWN is encountering higher density material and does not grow as quickly.
As a result of the combined effect of the smaller PWN radius and its motion towards the southeast, the reverse shock does not reach the northwestern hemisphere of the PWN for higher ejecta masses, and, therefore, a fraction of the ejecta remain unshocked at the final age of 2500~years. This is most evident in the bottom panel showing the density structure for Run~5 with an ejecta mass of 5.0$\rm \: M_{\odot}$. However, \citet{bhalerao19} find no evidence for a lower shocked ejecta mass in the northwestern hemisphere with respect to the southeastern hemisphere, and in fact find a higher shocked mass in the NW by a factor of $\sim$1.4. This suggests that the reverse shock has either shocked all ejecta or the majority of the ejecta in the NW, favoring the models with lower ejecta masses in the top and middle panels.

There are also some morphological differences between the runs shown in Figure~\ref{hydro_2panel}. The shape of the PWN for the model with $M_{ej}=5\rm \: M_{\odot}$ is  less circular than for the lower mass models and the displacement of the pulsar from the geometric center of the PWN is much lower in comparison. The observed pulsar position is approximately one third of the diameter away from the southeastern edge of the PWN (see Figure~\ref{radio_ir}, similar to runs 1 and 3 in top and bottom panels of Figure~\ref{hydro_2panel}.
These differences in PWN morphology and pulsar position therefore also favor the lower ejecta mass models. Based on the HD model results and their comparison with observations, we conclude that the reverse shock has either encountered the entire surface of the PWN or is close to the PWN boundary in the NW. This requires the ejecta mass to be close to the lower-mass range of the ejecta masses we explored, likely less than 2.5$\rm \: M_{\odot}$.

Lastly, in Figure~\ref{synch_life}, we show the age of the particles injected by the pulsar at the PWN age of 2500~years for the HD simulation for Run~1. For a magnetic field of $\sim$13~$\mu$G (see Table~\ref{model_param}), we estimate the synchrotron lifetime of 3~keV-emitting particles to be $\sim$450~years. The particles younger than this lifetime are roughly contained within the blue-color region in the left panel of Figure~\ref{synch_life}. These younger particles are expected to give rise to X-ray emission, while the older particles shown in red would primarily emit below the X-ray band. The right panel of the figure shows the \textit{Chandra} X-ray image in the 3--6~keV band, showing the extent of the PWN. The radio contours are overlaid in white \citep{gaensler03} and are seen to extend beyond the X-ray nebula. This is approximately consistent with the sizes expected from the HD simulation in the left panel. We note that all three HD runs shown in Figure~\ref{hydro_grid} produce a similar effect.

\subsection{Evolution into a Wind Density Profile} \label{wind_profile}

\begin{figure}
\includegraphics[width=0.47\textwidth]{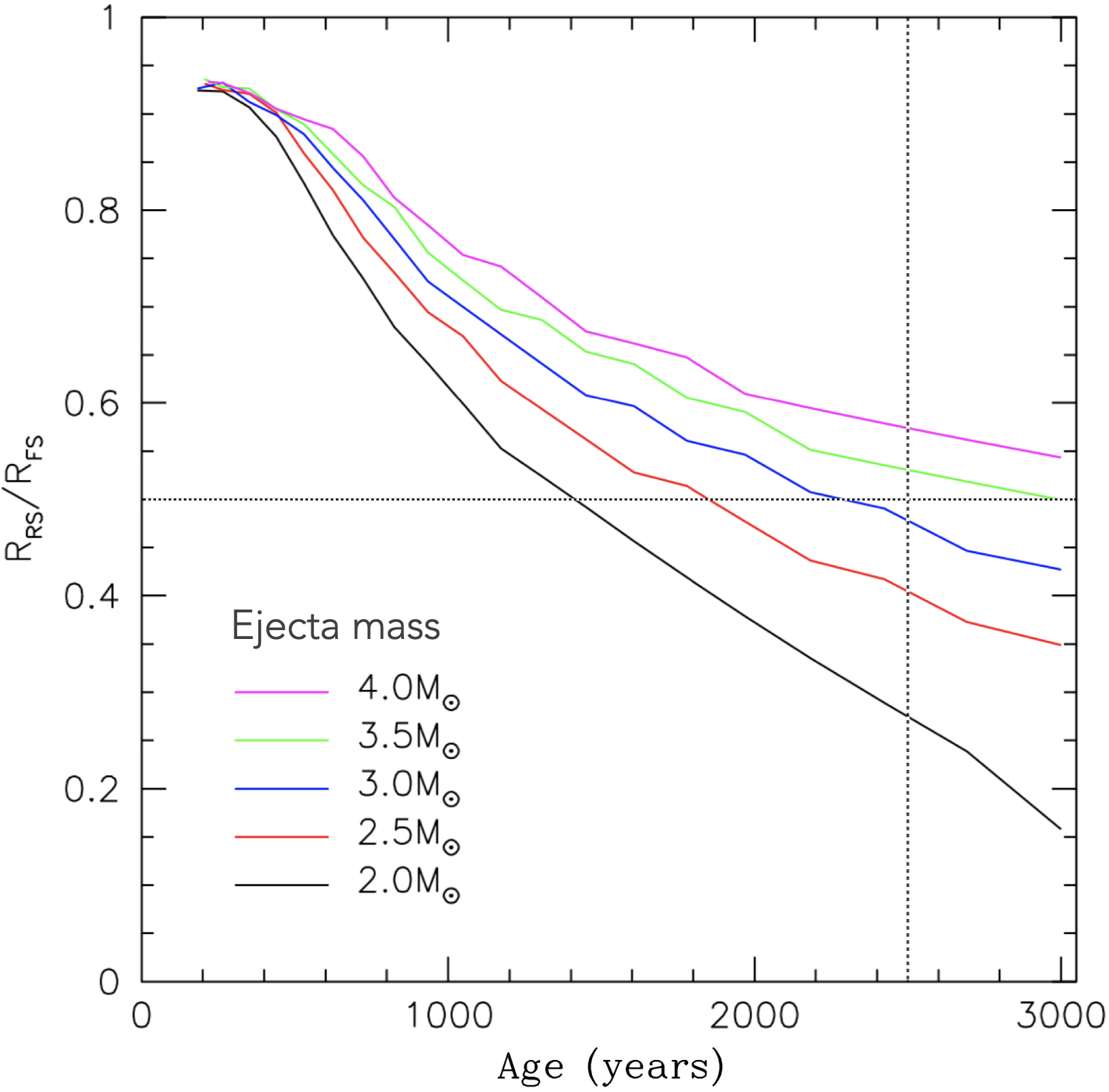}
\caption{\label{windprofile} Results of 1D hydrodynamic simulations of the evolution of the reverse-to-forward shock ratio ($\rm R_{RS}/R_{FS}$) as a function of SNR age for a radially decreasing ambient density profile of the form $\rho=\dot{M}/4\pi r^2v_w$, where the moss loss rate $\dot{M}=2\times10^{-5}\:\rm M_{\odot}\:yr^{-1}$ and the wind velocity $v_w$=10~$\rm km\:s^{-1}$. The curves depict models with the same ambient density profile and explosion energy of $\rm 10^{51}\: erg$, but different ejecta masses ranging from 2 to 4~$\rm M_{\odot}$. The dotted horizontal and vertical lines represent the $\rm R_{RS}/R_{FS}$ and SNR age for G292.0+1.8.}
\end{figure}

The dynamical models of the PWN evolution inside an SNR described in sections \ref{1d} and \ref{hydro} assume that the SNR expands into a uniform density medium. \citet{lee10} found that that X-ray observations of G292.0+1.8 are consistent with a radially decreasing ambient density profile with a density of $n_H=0.1-0.3\:$cm$^{-3}$ at the current SNR radius of 7.7~pc. Here, we employ a one-dimensional hydrodynamic model to explore the effects of a radially decreasing ambient density profile on the evolution of the SNR and the total ejecta mass that it contains. The one-dimensional code is based on VH-1, a multidimensional hydrodynamics code which has been modified to follow the thermal and dynamical evolution of supernovae and supernova remnants \citep{blondin93,patnaude15}. Figure~\ref{windprofile} shows the evolution of the reverse-to-forward shock radius ratio ($\rm R_{RS}/R_{FS}$) as a function of the SNR age for ejecta masses varying from 2--4~$ M_{\odot}$. All models have the same explosion energy of $10^{51}\:$ erg, the same radial wind density profile of the form $\rho=\dot{M}/4\pi r^2v_w$, where the mass loss rate $\dot{M}$ and wind velocity $v_w$ are assumed to be $2\times10^{-5}\: M_{\odot}\:$yr$^{-1}$ and 10~km$\:$s$^{-1}$, respectively. These parameters lead to an ambient density of $\sim0.1\:cm^{-3}$ at the current radius, in the range measured by \citet{lee10}. 

All models in Figure~\ref{windprofile} produce a forward shock radius within 10\% of the observed radius of 7.7~pc at an age of $\sim$~2500~years. Consequently, all models have approximately the same swept-up CSM mass of $\sim10\:M_{\odot}$ at that same age, increasing to 12.5~$M_{\odot}$ at 3000~years, a range consistent with the measured CSM mass in G292.0+1.8 \citep{bhalerao19}. The curves in Figure~\ref{windprofile} show that $\rm R_{RS}/R_{FS}$ decreases much quicker for lower ejecta masses, as may be expected. In previous sections, we concluded that the reverse shock in G292.0+1.8 is interacting with the PWN in the SE, requiring that the $\rm R_{RS}/R_{FS}$=0.5 at the present age. This ratio is indicated by the horizontal dotted line in Figure~\ref{windprofile}. At an age of 2500~years, indicated by the vertical dotted line, the observed reverse-to-forward shock ratio is consistent with an ejecta mass of $\sim\: 3\: M_{\odot}$, corresponding to a ZAMS progenitor mass of 14.6~$ M_{\odot}$ (derived by adding $\rm M_{ej}+M_{CSM}+M_{NS}$). The constant ambient density model Run 1 in Table~\ref{1Dparam} yields the same swept-up CSM mass and $\rm R_{RS}/R_{FS}$, but requires half of the ejecta mass. More broadly, the models in Figure~\ref{windprofile} show that SNR expansion into a radially decreasing density profile can yield $\rm R_{RS}/R_{FS}$=0.5 with a higher ejecta mass compared to the uniform ambient density models, while keeping the total swept-up ambient mass between the models the same and constrained to the measured mass range. However, the corresponding ZAMS mass is still relatively low and the ejecta-to-CSM mass ratios suggestive of a binary-stripped progenitor.

\section{Comparison with Nucleosynthesis Models}\label{abundmodel}

The key conclusion from the models discussed thus far is that the reverse shock in G292.0+1.8 is almost certainly interacting with the PWN in the southeast. This establishes the current location of the reverse shock and leads to a low ejecta mass ($\lesssim~3\: M_{\odot}$) being favored in the dynamical evolution of the PWN/SNR system. We also concluded that the reverse shock has most likely encountered the entire surface of the PWN at the present age or is very close to the PWN boundary; the same conclusion was made by \citet{bhalerao15} based on X-ray observations and \citet{gaensler03} based on radio observations. The important implication of this is that all or most of the ejecta have been reverse-shocked. With this in mind, we are assuming that the total elemental masses and abundances derived from X-ray observations \citep{bhalerao19} represent most of the ejecta, with no significant hidden unshocked material inside of the reverse shock. As will be discussed in later sections, we argue that the mass of material emitting at optical and IR wavelengths is very low compared to the X-ray emitting mass. In this section, we will compare the observationally estimated elemental masses to neucleosynthesis models of \citet{jacovich21} and \citet{sukhbold16}. Our treatment differs from previous studies on G292.0+1.8 that are summarized in Section \ref{ejecta} because we are comparing the total elemental masses to necleosynthesis models in addition to their abundances ratios.

\begin{figure*}
\center
\includegraphics[width=0.75\textwidth]{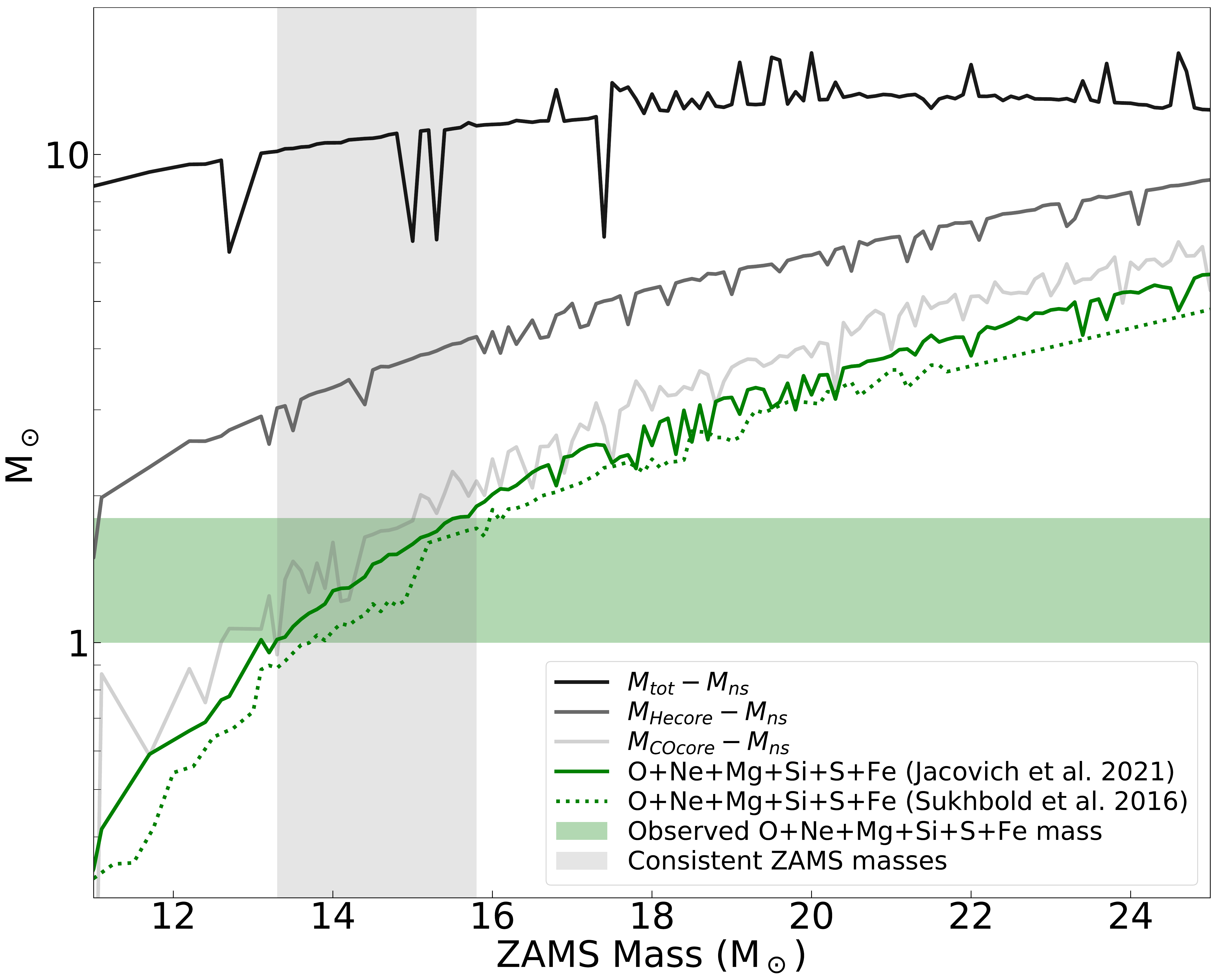}
\caption{\label{co_core_plot} The total ejecta mass (black curve), He core mass (dark gray curve), and CO core mass (light gray curve) as a function of the Zero Age Main Sequence (ZAMS) mass for the single star core-collapse models of \citet{jacovich21}. The green curve represents the total mass of O, Ne, Mg, Si, S, and Fe inside the He core as a function of ZAMS mass. All curves exclude the  neutron star mass. The horizontal green band indicates the total X-ray-measured mass of O, Ne, Mg, Si, S, and Fe, including the uncertainties \citep{bhalerao19}. The gray vertical band marks the corresponding range of consistent ZAMS masses.}
\end{figure*}

\subsection{Model Caveats} \label{caveats}

The core-collapse SNR models described in \citet{jacovich21} were derived for ZAMS masses of 10--30$\rm \: M_{\odot}$ and were evolved from the pre-main sequence stage, including wind-driven mass-loss.
The models involve \texttt{SNEC} \citep{morozova:15} simulations that make use of the \texttt{approx21} network \citep{timmes00}. An important consideration in using a limited network of isotopes is the amount of material that is
burned in a particular regime of $T$--$\rho$ parameter space. For instance,
the photodisintegration $^{20}$Ne($\gamma, \alpha$)$^{16}$O reaction will
enhance the production of $^{24}$Mg by a factor of $\lesssim$ 10 as compared
to reaction networks with larger numbers of isotopes. The reason behind this
is that in more advanced networks, neutrino cooling in the burning layers
can lead to increased densities, which will prevent vigorous Ne ignition
\citep{farmer16}.

Another important caveat of the \texttt{approx21} network is the inclusion
of a ``neutron-eating'' isotope of chromium, which becomes relevant during
$^{28}$Si burning. Due to the large Coulomb barrier, $^{28}$Si does not
generally burn to $^{56}$Ni. Instead, $^{28}$Si is first photodisintigrated,
and then a chain of reactions in quasi-statistical equilibrium (QSE) results in
the formation of the Fe-group elements \citep{hix96}.  QSE reactions often
result in a high degree of neutron-rich isotopes. However, with the inclusion
of the neutron-rich isotope of chromium, these neutron-rich isotopes of
other Fe-group elements are absent. 

In practical terms, all of these
effects impact the final ejecta composition profile, leading to an
overprediction in the Mg yield and an underprediction for Ne and some
Fe-group elements. Furthermore, it is important to note that the models
do not account for electron capture rates, and therefore may include
systemic errors in the abundances of the lowest mass progenitors and
in the Si core density profile because of the impact of electron
degeneracy pressure in this region. One final caveat of the models is that the
explosive burning in \texttt{SNEC} is sensitive to input parameters such as the neutron star mass cut, chosen explosion energy and duration, etc. The models in \citet{jacovich21} were tuned to broadly match results from \citet{young06}, but \citet{jacovich21} do note the sensitivity of the final yields to the chosen input parameters.

\begin{figure*}
\center
\includegraphics[width=1.0\textwidth]{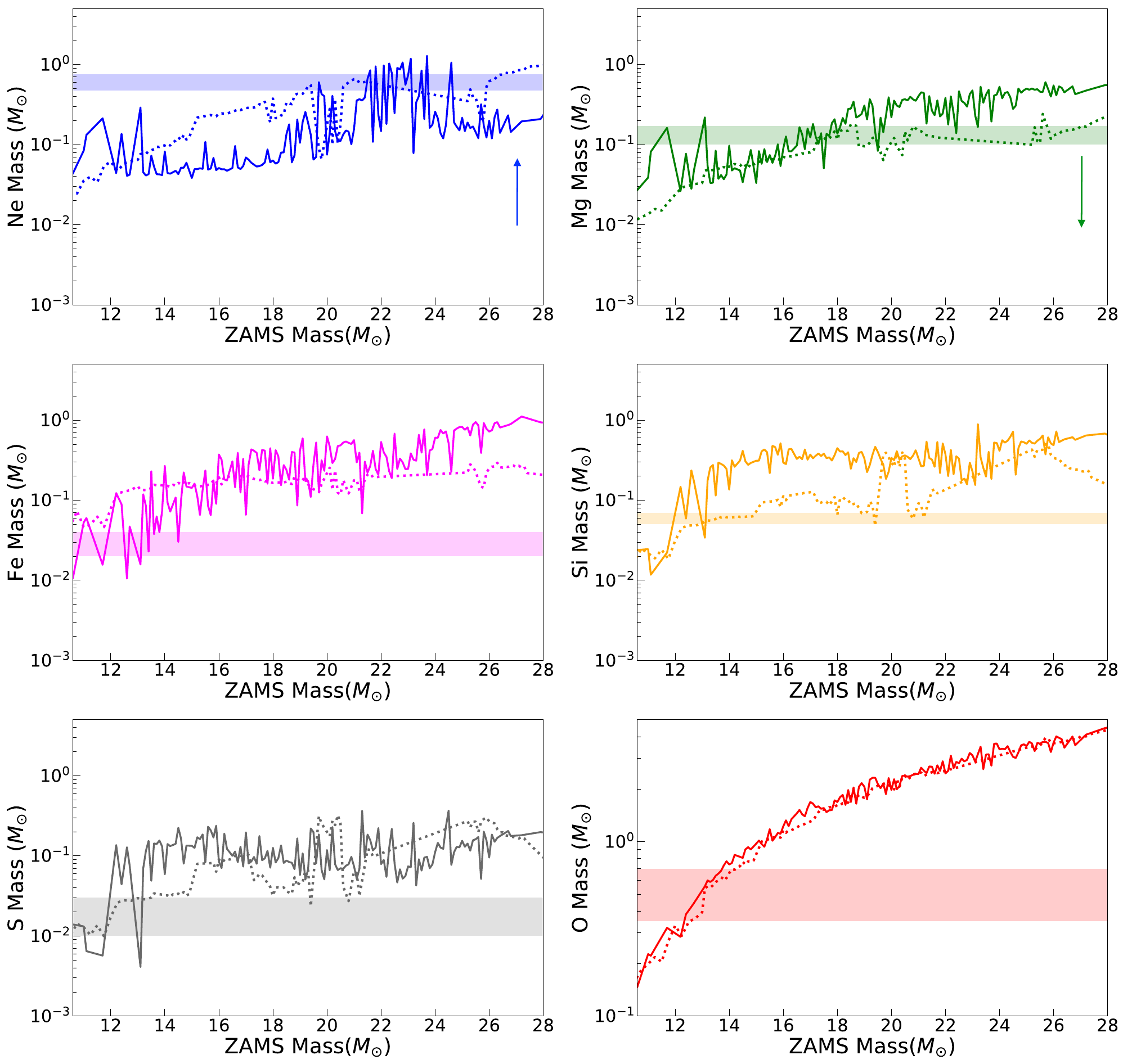}
\caption{\label{co_core_abund}The curves represent the total elemental masses (excluding the neutron star) as a function of the ZAMS mass, based on the core-collapse models of \citet{jacovich21} (solid lines) and \citet{sukhbold16} (dotted lines). We note that these models overpredict Mg and underpredict Ne masses (see Section~\ref{caveats}), with the colored arrows indicating in which direction the curves would be expected to shift in a more realistic scenario. The colored bands represent the elemental masses measured from X-ray observations of G292.0+1.8 by \citet{bhalerao19}.
}
\end{figure*}

\begin{figure*}
\center
\includegraphics[width=1.0\textwidth]{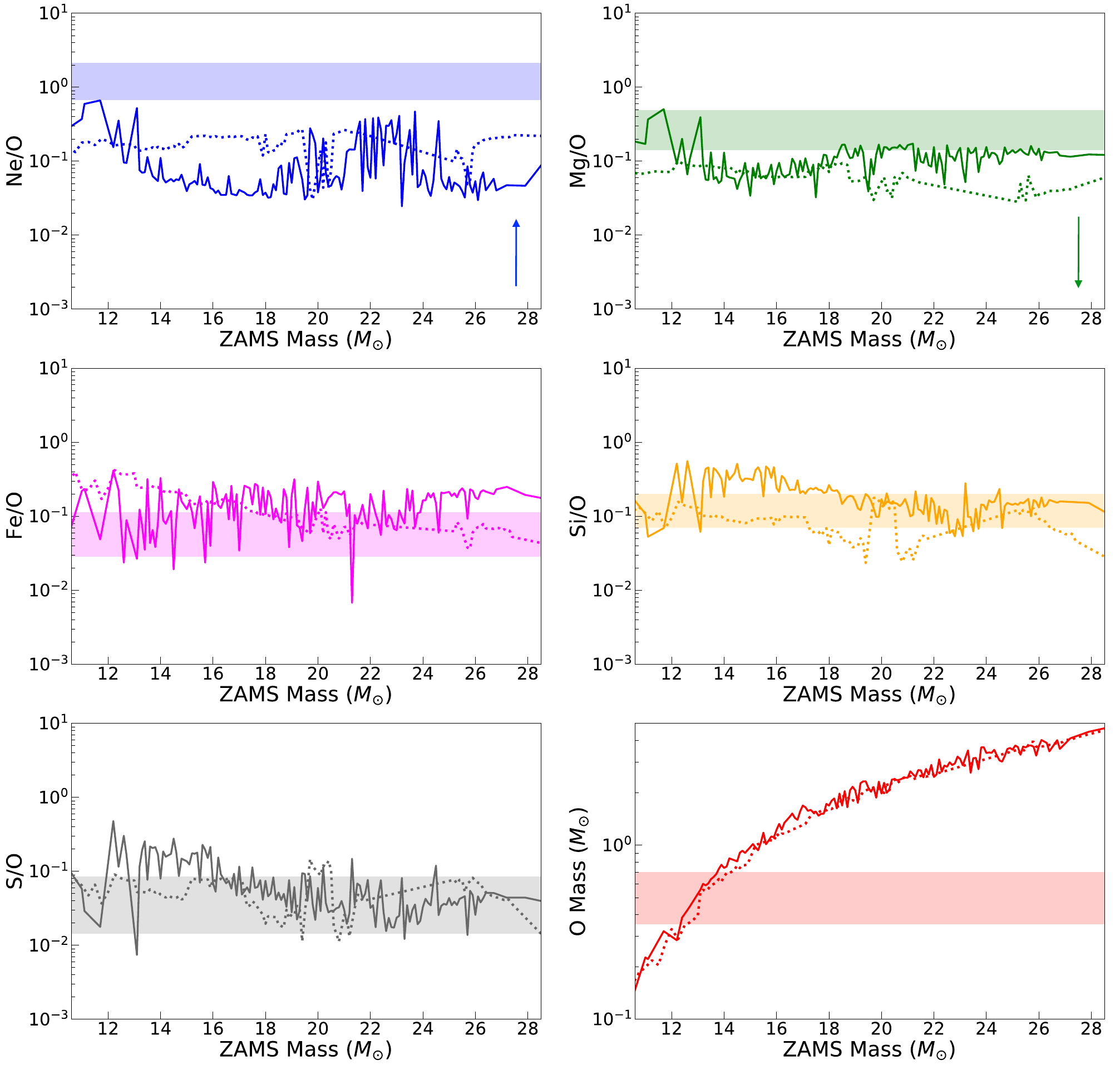}
\caption{\label{model_abund}The curves represent the elemental mass ratios (after the subtraction of the neutron star mass) as a function of the ZAMS mass, based on the core-collapse models of \citet{jacovich21} (solid lines) and \citet{sukhbold16} (dotted lines). We note that since these models overpredict Mg and underpredict Ne masses (see Section~\ref{caveats}), the Mg/O curve would move down and Ne/O would move up, as indicated by the colored arrows. The colored bands represent the elemental mass ratios as measured from X-ray observations by \citet{bhalerao19}. A ZAMS mass of around 13$M_\odot$ gives abundances ratios that are roughly consistent with total oxygen mass in the last panel and the measured ratios of Fe, Si, and S with respect to oxygen.}
\end{figure*}

\subsection{Comparison with Observed Abundances in G292.0+1.8} \label{abund}

The green curve in Figure~\ref{co_core_plot} shows the total mass of O, Ne, Mg, Si, S, and Fe (including $^{56} \rm Ni$) as a function of ZAMS mass for the models of \citet{jacovich21} (solid lines) and \citet{sukhbold16} (dotted lines). The curves exclude the neutron star mass. Since the majority of the mass of these elements is synthesized within the CO core, their total mass does not deviate significantly from the green curve as a function of how far out we integrate the radial abundance profiles.
The green band is the total mass of the same set of elements measured for G292.0+1.8 from X-ray observations \citep{bhalerao19}. The width of the green band reflects the uncertainties in the measured mass. 
The gray vertical band represents the corresponding ZAMS masses that are consistent with the observations, assuming that all of the ejecta in the SNR are reverse-shocked and emitting in X-rays. This range is roughly between 13.5 and 16~$\rm \: M_{\odot}$. For comparison, Figure~\ref{co_core_plot} also shows the CO and He core masses as light and dark gray curves, respectively, with the neutron star mass subtracted. The comparison of these curves to the ejecta masses of $\lesssim~3\: M_{\odot}$ favored by the dynamical models in Sections~\ref{1d} and \ref{hydro} suggests a highly stripped progenitor. The black curve shows the total ejecta mass as a function of ZAMS mass for the same single-star models using mass loss prescriptions described in \citet{jacovich21}. It can be seen that these masses greatly exceed the ejecta mass range favored by our dynamical models.

Next, we look at the constraints on the ZAMS mass from the individual element masses measured by \citet{bhalerao19}. The panels in Figure~\ref{co_core_abund} show the total mass of each observed element (excluding the neutron star mass) as a function of ZAMS mass for the yields from \citet{jacovich21} (solid lines) and \citet{sukhbold16} (dotted lines).
The colored bands are the X-ray-measured masses, with the band's thickness representing the uncertainty range. The total measured masses of Fe, Si, S, and O are consistent with a ZAMS mass in the 11--14~$\rm \: M_{\odot}$ range. Since Ne is underpredicted and Mg overpredicted in the models, the curves depicting the total mass of Ne (Mg)  would shift up (down) in the plots in Figure~\ref{co_core_abund}. While the total mass of Ne could come closer to the observed value for lower ZAMS masses, the mass of Mg will likely be more consistent with higher ZAMS masses, except for some ZAMS masses between 11 and 13~$\rm \: M_{\odot}$ in the \citet{jacovich21} models.

Figure~\ref{model_abund} shows the abundance mass ratios with respect to oxygen for the same models with the last panel showing the total oxygen mass.
We can see that the Fe, Si, and S ratios with respect to oxygen are consistent with observed ratios for a range of different ZAMS masses and are not particularly constraining.  The modeled Ne/O abundance is on the low end of the measured range for the lowest progenitor masses and too low for ZAMS masses above 12~$\rm \: M_{\odot}$. The higher (lower) masses of Ne (Mg) that would be expected for a more complete reaction network would actually bring their ratios in Figure~\ref{model_abund} closer to values consistent with the 12-14~$\rm \: M_{\odot}$ progenitor mass range.
Despite the many uncertainties in the models discussed in the previous section, it does appear that lower ZAMS progenitors are more consistent with the total observed elemental masses presented in this section. The abundance ratios are consistent with a wider range of ZAMS masses, including the lower mass progenitors.
If the progenitor of G292.0+1.8 is indeed in the 13--14~$\rm \: M_{\odot}$ range, the low Fe mass observed in the SNR may be less of an anomaly and within the range of what may be expected.  Combined with the low ejecta mass favored by the dynamical models, this conclusion would also imply that the progenitor experienced significant mass loss through binary stripping.

\section{The Nature of the Spur} \label{spur_nature}

\input{tab3.tex}


\subsection{Emission Properties}

The first detailed kinematic study of G292.0+1.8 subsequent to the discovery of optical emission by \cite{murdin79} was performed by \cite{ghavamian05}, who used Fabry-Perot imaging spectrometry in [O~III] $\lambda$5007 to map the radial velocities of the Spur.  Among
the findings of this study were the following: (1) radial velocities extending from 0 to +1500 km/s in the Spur, with no evidence of any blueshfited emission, (2) an unusual kinematic trend wherein the top of the Spur has negligible Doppler shift, while the lower portions of the Spur show increasingly redshifted emission.   The velocity pattern cannot be explained with a simple expanding shell model (i.e., an ellipsoid in position-velocity).  In their deep Spitzer IRS observations of the spur, \cite{ghavamian09} detected [\ion{O}{4}]~25.8~\micron, [\ion{Ne}{3}]~15.56~\micron\/ and  [\ion{Ne}{5}]~24.32~\micron.   However, they saw no firm evidence for any magnesium, argon, sulfur or iron line emission.  Subsequent observations confirmed these results \citep{ghavamian12}, reinforcing the impression from optical studies \citep{ghavamian05} that very little heavy element ejecta beyond those expected from the hydrostatic layers of the progenitor star were present in the Spur.

In the right panel of Figure~\ref{radio_ir}, we show the \ion{O}{4} 26~\micron\ emission obtained from the \textit{Spitzer} IRS spectral cube originally published by \citet{ghavamian12}. The radio contours outlining the radio PWN and SNR shell are overlaid. The line emission maps of other lines are shown in Figure~\ref{ir_velocities}. As also noted by \cite{ghavamian12}, the spatial morphology of \ion{Ne}{3} closely traces the emission from \ion{O}{4}. Emission from \ion{Ne}{5} is detected at the location of the brightest peaks in the \ion{O}{4} and \ion{Ne}{3} images. The morphology of the \ion{Si}{2} emission does not resemble the general morphology of the oxygen and neon lines. The \ion{S}{3} emission is very faint and also does not show any obvious correlation with the other lines.

We have measured the line centroid shifts of the [\ion{O}{4}]~25.89~\micron, [\ion{Ne}{3}]~15.56~\micron, [\ion{Si}{2}]~34.82~\micron, [\ion{Ne}{5}]~24.32~\micron, and [\ion{S}{3}]~18.71~\micron\ lines extracted from 30\arcsec $\times$ 30\arcsec\ square apertures arranged in a grid across the PWN. The corresponding velocity shifts with the uncertainties in parentheses are plotted on the line emission maps. We note that the blue and redshifted line components are not spectrally resolved in the IRS spectra and that the measured lines are the blend of these two components. Any blue or red shifts in the centroids are, therefore, likely a lower limit on the shift. It is evident from the \ion{O}{4} and \ion{Ne}{3} maps in Figure~\ref{ir_velocities} that the emission lines in the spur are highly red-shifted, as observed by \citep{ghavamian05}. The difference in the velocity distribution of the \ion{Si}{2} lines suggests that it does not arise from the same material, except in one area within the PWN for which the emission is redshifted. The \ion{S}{3} line was detected in only one extraction region, as can been seen in the last panel of the figure.

\subsection{Association with the PWN} \label{spur}

Based on the morphology and dynamics of the emission from the spur, we argue that this structure is associated with the PWN in G292.0+1.8.The spatial morphology of the oxygen emission is correlated with the PWN morphology. In the east, the emission appears contained within the PWN. The emission from oxygen and neon is also concentrated on the side of the PWN that coincides with the pulsar's direction of motion, as indicated by the black arrows in Figure~\ref{ir_velocities}. Furthermore, the highest velocity redshift in the emission lines occurs roughly in the direction of the pulsar's motion. This type of velocity and emission structure could be explained if the pulsar also has a line-of-sight velocity component away from the observer. The PWN would have shocked the highest velocity ejecta material in the direction of the pulsar's motion compared to the rest of PWN's surface, and this material would primarily be redshifted. \citet{ghavamian05} showed that optical emission lines in the spur structure span a velocity range between -720 and 1440~$\rm km\:s^{-1}$. If we assume that the SNR has a systemic velocity of 0~$\rm km\:s^{-1}$ \citep{gaensler03}, and that the ejecta expand symmetrically, this would imply a line-of-sight PWN velocity of $\sim$~360~$\rm km\:s^{-1}$, giving a total space velocity of approximately 600~$\rm km\:s^{-1}$.

The HD simulations discussed in Section~\ref{hydro} give some clues about the origin of the spur emission. The top panel of Figure~\ref{hydro_2panel} shows the density structure for Run~1 at the final age of 2500~years. It can be seen that the thin red shell of ejecta material surrounding the PWN is thicker, denser, and less stable in the southeastern side of the PWN. Prior to the reverse shock collision, the pulsar's and PWN's motion through the SN ejecta would have resulted in a significantly higher shock velocity in this direction.
The emission from the spur may arise from this gas at the PWN boundary. Another possibility is that the emission may originate from material that has been shocked by the reflected shock that propagates back into the large-scale Rayleigh-Taylor filaments following the collision of the reverse shock with the PWN boundary in the southeast. The propagation of this shock can be seen in the bottom panels of Figure~\ref{hydro_grid}. The IR and optical emission would arise from dense filament cores encountered by the reflected shock, and if the pulsar's velocity has a line-of-sight component, the emission would appear to be within the PWN in projection. This scenario could also explain the IR emission observed beyond the radio contours of the PWN in the south, most evident in the right panel of Figure~\ref{radio_ir}. Alternatively, the reflected shock or the PWN could be encountering the bi-polar structure of high-velocity knots observed at optical wavelengths \citep{winkler09}, giving rise to enhanced IR emission in the direction of the pulsar's motion.

\subsection{Shock Models}

Next, we explore the abundances in the spur to determine if they are consistent with the X-ray measured abundance ratios. Since we concluded that all the SN ejecta material has been reached by the reverse shock, we may expect that the abundances in the IR and optical should be similar to those found in X-rays.

\begin{figure*}
\center
\includegraphics[width=0.8\textwidth]{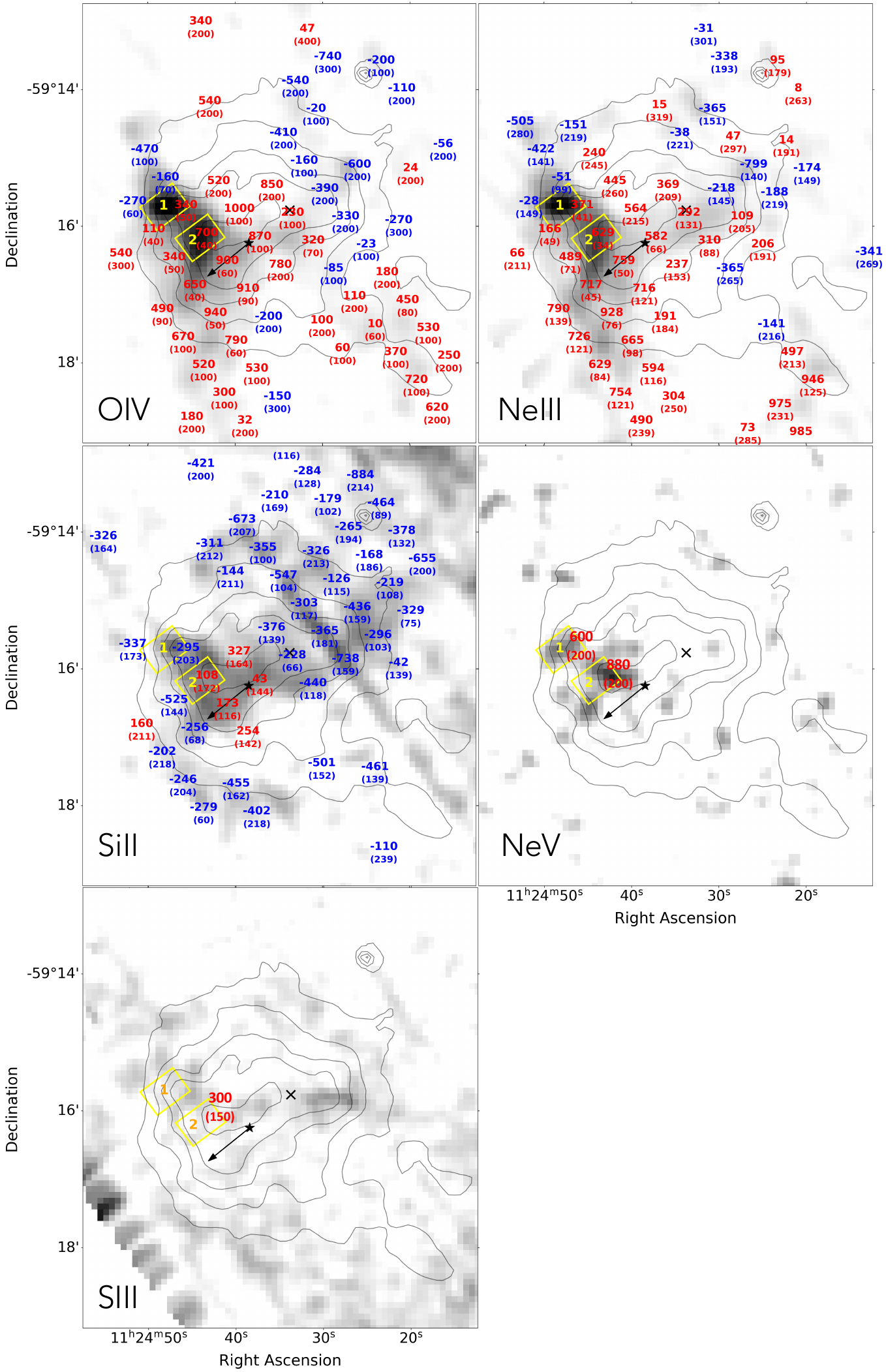}
\caption{\label{ir_velocities}Narrow band images of the [\ion{O}{4}]~25.89~\micron, [\ion{Ne}{3}]~15.56~\micron, [\ion{Si}{2}]~34.82~\micron, [\ion{Ne}{5}]~24.32~\micron, and [\ion{S}{3}]~18.71~\micron\ lines produced from the \textit{Spitzer} IRS spectral cubes (data originally presented in \citet{ghavamian12}). The black radio contours from the PWN are overlaid. Velocity shifts of the emission lines and their uncertainties are plotted on the images (red-shifted centroids in red and blue-shifted ones in blue). The yellow rectangles are the extraction regions for the spectral line measurements listed in Table~\ref{ir_lines}.}
\end{figure*}

We applied the shock models to measured emission line properties extracted from two positions on the spur, listed in Table~\ref{ir_lines}. The extraction regions are indicated by the yellow rectangles in Figure~\ref{ir_velocities} and their positions roughly correspond to those analyzed by \citet{ghavamian09}. We also included the optical line measurements for those positions provided by \citet{ghavamian09}. We note that additional uncertainties arise from modeling the IR line ratios since the blue- and red-shifted line components are not resolved in the IRS spectra.

\subsubsection{Background and Caveats}

Several models of the emission spectra from shocks in SN ejecta have been made in order to interpret the emission spectra \citep{raymond18}.  \citet{itoh81a, itoh81b} modeled shock waves in pure oxygen, emphasizing the difference between ion and electron temperatures that comes about when the electron cooling rate due to excitation of ions exceeds the ion-electron energy transfer rate due to Coulomb collisions.  \citet{borkowski90} emphasized the importance of thermal conduction, and \citet{dopita84} showed that mixtures of elements can produce much different cooling rates.  \citet{sutherland95} studied the importance of emission from the photoionization precursor, and \citet{blair00} emphasized the importance of UV emission that is observable in oxygen-rich remnants such as N132D and 1E1021.2-7219, but not in the highly reddened Galactic remnants Cas A and G292.   \citet{docenko10} used models that include the photoionization precursor, but not electron-ion temperature differences or thermal conduction, to analyze IR observations of the knots in Cas A.  \citet{koo13} and \citet{koo16} emphasized the near IR lines of [Fe II] and [P II].

There are basic uncertainties about the physics that goes into these models, including the suppression of thermal conduction by magnetic fields, the effects of photoionization by X-rays from hotter parts of the SNR or from the PWN, and the degree of electron-ion equilibration at the shock (though even if T$_e$=T$_i$ at the shock, T$_e$ rapidly falls below T$_i$ because of the strong cooling in shocks slower than a few hundred $\rm km~s^{-1}$).  There is also a persistent problem that the models overpredict the O I recombination line at $\lambda$7774 unless the cooling flow is arbitrarily cut off near 1000 K \citep{itoh86}.  A second problem is that the gas tends to remain at a nearly constant electron temperature while it is heated by Coulomb collisions with the ions.  Then, the temperature plunges quickly when the ions cool off.  That results in, for instance, a shock that produces strong [O III] but practically no [O II] or [O I] emission.  

The fundamental problem is probably that the models assume planar geometry and steady flow, neither of which is likely to be a justified approximation for a shock that encounters a dense cloud.  The cloud will experience different shock speeds at its nose and on its flanks, and it will be rapidly shredded by shear instabilities \citep{klein94}. \citet{eriksen09} presented a multi-dimensional model of the shock-cloud interaction, including time-dependent ionization.  However, he assumed a pure oxygen composition, and the spatial resolution of the code limited the accuracy of any spectral predictions.

\subsubsection{Application to the Spur} \label{shock_app}

Given the difficulties with shock models described above and the small number of observable lines, we make the following assumptions.

-- We assume, following \citet{docenko10}, that the [Ne II] emission comes from the photoionization precursor.

-- Because the ratio of the two [Ne V] lines depends on density, with a critical density of about $3 \times 10^4~\rm~cm^{-3}$, and the observed ratio observed at position 1 matches the low density limit according to the CHIANTI package \citep{delzanna15}, we assume that the preshock density is of order 1~$\rm cm^{-3}$, though higher densities are possible if the magnetic field halts the compression of the cooling gas.  The [Ne V] ratio at position 2 suggests a higher density, but the lines are relatively weak, and therefore uncertain.

-- The PWN emission has little effect, even on the low density preshock gas.  The PWN luminosity is about $2 \times 10^{35}~\rm erg~s^{-1}$, which for a density of 1~$\rm cm^{-3}$ at a distance of 3.8 pc from the center, gives an ionization parameter $\xi = L/(n r^2)$ of only $10^{-3}$.

-- Because of the model uncertainties, we employ a model that matches the ionization state, in particular the [Ne III]/[Ne V] ratio at position 1, and we assume that it correctly gives the relative ionization of [O IV], which is intermediate between [Ne III] and [Ne V] in either collisionally ionized or photoionized plasmas. This model has a shock speed of 125 $\rm km~s^{-1}$ at positon 1 and perhaps 100 $\rm km~s^{-1}$ at position 2, but those values should be considered to be quite uncertain.

Table \ref{shock_table} compares the observed [Ne III] and [Ne V] to [O IV] ratios with models similar to those of \citet{blair00} with the average G292.0+1.8 abundance set of \citet{bhalerao19} derived from X-rays (O:Ne:Mg:Si:S:Fe = 0.47:0.57:0.13:0.06:0.02:0.03 by mass).  The IR and optical lines are normalized to [O IV] = 100 or [O III] = 100, respectively, and the optical lines are taken from \citet{ghavamian09}. 
The model assumes nearly complete ion-electron temperature equilibration at the shock, but the electrons become much cooler than the ions just behind the shock, and they regain equilibrium only when the gas cools to around 10$^4$ K, where the Coulomb collision rate is higher and the relevant collisional excitation rates are lower.  Illumination by the PWN is included, but it makes little difference.  The magnetic field is assumed to be weak, since the ejecta have expanded by a large factor.  However, the field could be significant if it is enhanced by turbulence in the cloud-shock interaction or if it was enhanced when the gas was shocked at an earlier time.

As mentioned in section 5.3.1, it is a longstanding problem that a model shock in highly enriched gas tends to produce strong emission in just a few ions and very little in lower ionization states, in contrast to shocks in gas with normal astrophysical abundances.  This is because the cooling rate is far higher than the recombination rates.  This problem is apparent in the gross disagreement between models and observations of the [O~II]/[O~III] ratio.  It is usually explained with a sum of shock models of different velocities to match the different ionization states, but we do not have enough observables to deal with the additional free parameters.  Therefore, in assessing the S abundance, we compare [S II] with [O~II] rather than [O III].  The observed [O II]/[S II] ratio is  18, with some uncertainty due to the reddening correction, while models based on the \citet{bhalerao19} abundance set predict around 11.  We consdier that to be within the uncertainties, thus supporting the idea that the \citet{bhalerao19} abundances pertain to the ejecta as a whole.   

\input{tab4.tex}

Since the [Ne III] to [Ne V] ratio is very sensitive to the assumed shock speed, we use ([Ne III] + [Ne V])/[O IV] to estimate that the Ne:O ratio is somewhat higher than the X-ray value of 1.6 by mass or 1.3 by number. For position 1, Ne:O = 2.1 by number would match the IR line ratios.  This is somewhat outside the uncertainties given by \citet{bhalerao19}.  However, the  optical [Ne III]:[O III] ratio for position 1 given by \citet{ghavamian09} suggests that the Ne:O ratio should be around 2/3 the X-ray abundance ratio, or about 0.9.  This might reflect an actual variation in the abundances, since the IR and optical observations of position 1 do not completely overlap.  Another possibility is that, while the IR lines are insensitive to the  temperature of the emitting region, the optical lines are sensitive to T, and the model temperatures might not be accurate. Either way, the shock model analysis of the IR emission in the spur confirms the overabundance of Ne in the ejecta material that has been observed at X-ray wavelengths.

\subsection{Mass of Dust and Gas in the Spur}

In order to estimate how much ejecta material may be contained in the spur, we assumed that the material is distributed in a spherical cap around the PWN with a base radius of 45~\arcsec\ and a thickness of 5~\arcsec, as approximately measured from the optical data of \citep{ghavamian05}. For a distance of 6~kpc, this gives a total volume of $\sim$~0.8~$\rm pc^{3}$. Assuming an upper limit on the density of 1~$\rm cm^{-3}$ (see Section~\ref{shock_app}) and an oxygen composition, we estimate that the spur does not contain more than 0.3~$\rm \: M_{\odot}$ of ejecta.
\citet{ghavamian16} find that the total dust mass in the spur to be only $1.2\times10^{-3}\rm \: M_{\odot}$, leading to a dust to gas ratio of $\sim$~0.004. Since the reverse shock has already collided with the PWN, it is interesting to note that the dust emission in the spur arises from grains that have survived the passage of the reverse shock.

\section{Summary \& Discussion} \label{summary}

\subsection{The Progenitor of G292.0+1.8} \label{progenitor}

In previous sections, we showed that the dynamical modeling of G292.0+1.8 and its PWN  favors a relatively low ejecta mass, on the order of
$\lesssim~3\: M_{\odot}$, and that the observed elemental masses
and ratios favor a relatively low ZAMS mass in the
12-16~$\rm \: M_{\odot}$ range. Combined, this is indicative of a
relatively low mass, binary-stripped or even ultra-stripped-envelope progenitor for G292.0+1.8 that would have resulted in a Type IIb, Ib or Ic SN explosion. Below
is a summary of the evidence that points towards this conclusion.

\begin{enumerate}[wide=0pt, listparindent=1.25em, parsep=0pt]

\item Based on the HD simulations described in Section~\ref{hydro},
  the pulsar's displacement from the PWN center and the radio PWN's
  morphology and position in the SNR provide strong evidence that the
  PWN has been interacting with the reverse shock in the southeast and
  that it has either already encountered or is close to encountering
  the entire surface of the PWN. This conclusion is consistent with
  X-ray and radio observations of G292.0+1.8
  \citep{gaensler03,bhalerao15}. The HD simulations show that
  for this to occur, the total ejecta mass needs to be low, on the
  order of 2.5~$\rm \: M_{\odot}$ or less. In a recent study,
  \cite{prentice19} analyzed 18 stripped-envelope SNe, and combined
  with results for other sources available in the literature
  \citep{drout11,lyman16,taddia15, taddia18}, found the mean ejecta
  masses for Type IIb and Ib SNe to be 2.7$\pm$1.0~$\rm \: M_{\odot}$
  and 2.2$\pm$0.9~$\rm \: M_{\odot}$, respectively, and that
  narrow-line Type Ic SNe have the largest spread in ejecta masses
  with a range from 1.2--11~$\rm \: M_{\odot}$. The observed mass
  range strengthened the evidence that the majority of
  stripped-envelope SNe arise from low mass progenitors. The ejecta
  mass that we favor for G292.0+1.8 is in this range.

\item The 1D models of the evolution of the SNR into a uniform ambient medium show that in order for the reverse shock to reach far enough into the
  interior to interact with the PWN, a low ejecta mass is required. Otherwise, the required ambient CSM density is implausibly high, resulting in a
  total swept up CSM mass much higher than the value of
  $13.5^{+1.7}_{-1.4}f^{1/2}d_6^{3/2}\rm \: M_{\odot}$ measured from
  X-ray observations \citep{bhalerao19}. The low dust mass of
  0.0023~$\rm \: M_{\odot}$ associated with the CSM emission
  \citep{lee09,ghavamian16} results in a dust-to-gas mass ratio
  several times lower than the Galactic average when the X-ray
  measured CSM mass is used, providing further evidence that the total
  swept up CSM mass is low. We note that the 1D model discussed in
  Section~\ref{1d} actually resulted in a swept-up CSM mass that is
  too high even for $M_{ej}=2.5\rm \: M_{\odot}$ (see
  Table~\ref{1dmodel}). For example, for a 13.5~$\rm \: M_{\odot}$
  progenitor and  $M_{ej}=2.5\rm \: M_{\odot}$, the swept-up CSM mass should be on
  the order of only $\sim$~10~$\rm \: M_{\odot}$. 
Invoking a radially decreasing ambient density profile mitigates this issue since it allows for a somewhat higher ejecta mass with a relatively low swept-up CSM mass. For example, to match the forward and reverse shock radii of G292.0+1.8, the decreasing density profile leads to an ejecta mass of 3~$M_{\odot}$ and a swept-up CSM mass of 10~$\rm \: M_{\odot}$, leading to a ZAMS mass of 14.6~$\rm \: M_{\odot}$, consistent with our conclusion of a binary-stripped progenitor.

\item From the HD simulations we concluded that the reverse shock has
  reached all or most of the ejecta and that there is no significant
  unshocked material hidden in the interior of the SNR. We estimate
  that optical and IR-emitting spur contain at most
  0.3~$\rm \: M_{\odot}$ of ejecta. If we then assume that the X-ray
  measured masses of O, Ne, Mg, Si, S, and Fe account for most of the
  ejecta, comparison to single-star nucleosynthesis models lead to a ZAMS mass in the range of
  12--16~$\rm \: M_{\odot}$. While the Ne abundance is still
  somewhat high in comparison to oxygen, the lower ZAMS mass range
  may explain the low Fe mass that has been inferred from observations
  and previously considered to be an anomaly. 

\item The morphology of features in G292.0+1.8 might also be more naturally explained by
a binary progenitor, especially if the explosion was from the
initially more massive star in the system, that is, if the delay between a
binary interaction and the explosion is only the remaining lifetime of
the more evolved star. Specifically,
non-conservative binary interactions can replenish the CSM, and the
ring-like structure observed might be the leftover from a circumbinary
disk hit by the SN blast-wave (similar to the explanation for the
central ring in SN1987A \citep{morris:07, morris:09}.

\end{enumerate}

There are important caveats in our results and conclusions that are
worth consideration and future exploration. The 1D and
2D HD models we use do not capture the complexity of more realistic ambient density profiles that may
result from stellar winds, eruptive mass loss, or binary
interactions. For instance, the presence of the dense circumstellar
ring in G292.0+1.8 would clearly impact the dynamics but is not
accounted for in the simulations. Furthermore, we do not consider
other effects that may influence the dynamics, such as asymmetries in
the explosion and the SN ejecta profile. 
These considerations, in addition to uncertainties in the SNR distance and
precise age, would likely affect our conclusions about the total ejecta and swept-up ambient masses.

As noted by \citet{bhalerao19}, there may be additional systemic uncertainties in the elemental mass measurements from X-ray observations associated with the assumptions about the emitting volume and the filling factor that characterizes the clumpiness of the ejecta. \citet{bhalerao19} assumed that the X-ray emission from the ejecta arises from knots and filaments in localized volumes, rather than gas distributed over the entire volume of the shell spanned by the contact discontinuity and PWN radii. While an assumption of a larger emitting volume would increase the ejecta mass estimates, a filling factor smaller than unity would offset the increase. While we cannot estimate how large the uncertainty on the mass would be, we do not expect that it will be significant enough to affect our conclusions. We note that \citet{bhalerao19} assumes that the X-ray emission from the diffuse CSM arises from a thick shell, so their measured CSM mass is more likely to be an upper limit since the filling factor may be less than unity. Nevertheless, the comparison of the measured mass to the swept-up CSM mass from our dynamical models and the requirement for the reverse shock to be interacting with the PWN still imply that the mass of the SN ejecta
needs to be relatively low to produce the observed properties of
G292.0+1.8.

Besides the limitations of the smaller reaction network discussed in
Section~\ref{caveats}, the nucleosynthetic yields of
\citet{jacovich21} and \citet{sukhbold16} are based on single star progenitor models that may
not be applicable to binary-stripped stars. Binary-stripped
donor stars likely have less massive cores for a given initial mass  \citep[e.g.][]{schneider21}
and composition structures that may differ significantly from those of
single stars. \citet{laplace21} recently showed that, unlike single
stars, the pre-supernova core composition profiles of binary-stripped
have an extended gradient of carbon, oxygen, and neon, that could lead
to systemically different SN yields \citep[also see][]{farmer21}.
However, these studies show that the mass of oxygen, which is
sensitive to the CO core mass, remains similar between single and
binary-stripped stars. We therefore argue that our conclusion about
the ZAMS range for the progenitor of G292.0+1.8 would still hold.

\subsection{Implications for the Binary Companion}

Based on the low ejecta mass, the explosion that produced G292.0+1.8
was likely of a relatively low mass star ($<16\,M_\odot$). To
explain the low ejecta mass, throughout its evolution, the progenitor
must have lost much more mass than what is expected from stellar winds
\citep[e.g.,][]{renzo:17}. This likely exposed He-rich (and possibly
even CO-rich) layers at pre-explosion. \citet{zapartas:21b} explored
the impact on various SN engines on models of single stars, and found
that stars massive enough to lose their H-rich envelope by their own
winds are difficult to explode. The presence of a binary companion,
which is a common occurrence for massive stars
\citep[e.g.,][]{sana:12}, might provide the means to shed a large amount
of mass, and make the core-structure more explodable
\citep{schneider:21, vartanyan:21}.
\cite{zapartas:17b} studied the variety of evolutionary paths
leading to a stripped envelope SN, and found that, at
$Z=0.014\simeq Z_\odot$ (cf.~Fig.~6 in \citealt{zapartas:17b}), in
about $60\%$ of the cases the presence of a companion is expected.

Based on \citet{zapartas:17b}, $\sim{}56\%$ of the systems
resulting in a H-free core-collapse exhibiting a main-sequence
companion, so this would be the most likely case for G292.0+1.8. The companion can remain bound to the NS at the explosion
($14_{-9}^{+22}\%$ of cases, \citealt{renzo:19}) and can be searched
for using the NS as a guide \citep[e.g.,][]{kochanek:19}. The binaries
surviving a SN explosion however are typically characterized by
relatively small natal kicks, which is in tension with the large
pulsar velocity inferred for G292.0+1.8 (see Section~\ref{kick}).
In $86^{+9}_{-22}\%$ of the cases \citep{renzo:19}, binaries are disrupted at the first
SN and eject the companion. The present day position of the companion
and its appearance depend on the kind of binary interaction that
removed the envelope of the exploding star: common envelope or Roche
lobe overflow (RLOF) stripping. By definition, common envelope results in tight pre-explosion orbit
and thus large ejection velocity, typically of the order of
$\sim{}100\,\mathrm{km\ s^{-1}}$ \citep{renzo:19} but up to
$\sim{}1200\,\mathrm{km\ s^{-1}}$ \citep{tauris:15, evans:20}. Conversely, if the stripping of the progenitor of G292.0+1.8 occurred
during the first RLOF phase, the main-sequence companion is expected
to be a slow ``walkaway'' in $\sim{}95\%$ of the cases
\citep{renzo:19}, with a typical velocity of only a few
$\mathrm{km\ s^{-1}}$. Given the predicted velocity of main-sequence companions (see
\citealt{renzo:19}), and the time since the explosion, one can define
a radius for companion searches (see \citealt{kerzendorf:19} for a
similar approach for Cassiopea A).\\

\subsection{Natal Kick of the Neutron Star} \label{kick}

In Section~\ref{spur}, we suggest that the velocity structure of the
spur region may be explained if the pulsar has a line-of-sight velocity
component of $\sim$~360~$\rm km\:s^{-1}$, bringing the total space
velocity to approximately 600~$\rm km\:s^{-1}$. We note that this
inferred velocity
should not be considered a proxy for the natal kick only. In case the
progenitor was orbiting a companion, especially if
post-common-envelope, the orbital velocity can reach a few
$100\,\mathrm{km\ s^{-1}}$ and should be added to the kick velocity to
obtain the present day velocity \citep[see
also][]{kalogera:96,tauris:98}.
Nevertheless, the high velocity inferred is statistically unlikely to
be reached without a significant natal kick. Based on the masses of
Galactic double NS, it has been proposed that stripped (and
ultra-stripped) progenitors lead to smaller cores, easier to explode
without large asymmetries and consequently result in smaller natal
kicks \citep[e.g.,][]{podsiadlowski:04, muller15}. A strong correlation has been proposed between the neutron star kick velocity and progenitor mass \citep[e.g.][]{eldridge16}, and between the kick velocity and $M_{ej}$ and explosions energy \citep[e.g.][]{janka17}. This would seem
in tension with the very low inferred ejecta mass and the large NS
velocity for G292.0+1.8 that was found to correlate with the observed ejecta asymmetries \citep{hollandashford17, katsuda18b}. However, \cite{schneider:21} recently
challenged the conclusion of low natal kicks based on detailed stellar structures and
semi-analytic explosion models. In 3D models of neutrino-driven core-collapse SNe, \citet{muller19} achieve kicks as high as 695~$\rm km\:s^{-1}$ for ultra-stripped-envelope progenitors with He core masses in the 2.8--3.5~$\rm M_{\odot}$ range (including the neutron star), which does seem to be consistent with G292.0+1.8. 
Furthermore, natal kicks are likely a
stochastic outcome of SN explosions \citep[e.g.][]{mandel20}, decreasing the tension between
individual NS velocity measurements and expected kick predictions.

\section{Conclusion}\label{conclusion}

We present 1-D dynamical evolution modeling of SNR G292.0+1.8 and its PWN, as well as 2-D HD models that account for the pulsar's motion through the SNR. The simulations show that the pulsar's displacement from the geometric center of the PWN can only be achieved if the SNR reverse shock has collided with the PWN in the southeast. Furthermore, the comparison of the simulations to the morphology of the PWN at present age indicate the reverse shock has either reached the entire surface of the PWN or is very close to its boundary even in the northwest. The resulting required ratio of forward and reverse shock radii, and the observed total mass of the CSM in the remnant, are consistent with an ejecta mass of $\lesssim3\: M_{\odot}$. 

Since the HD modeling implies that the majority of the ejecta have been reverse-shocked, we compare the total elemental masses measured from X-ray observations \citep{bhalerao15} to nucleosynthesis yields and conclude that the total masses and abundance ratios favor a ZAMS mass in the 12--16~$\rm M_{\odot}$ range. A progenitor in this mass range may explain the low Fe mass in G292.0+1.8, but the observed Ne yields are still higher than models predict. Combined, our results point to a binary-stripped, perhaps even ultra-stripped, progenitor with a relatively low mass that might have produced a Type IIb or Ib/Ic SN.

We also conclude that optical- and IR-emitting ``spur" region in G292.1+1.8 arises as a result of the PWN's interaction with SN ejecta in the direction of the pulsar's motion. The velocity structure of the spur may indicate a line-of-sight velocity component to the pulsar's motion, resulting in a total space velocity of $\sim$600~${\rm\ km\: s^{-1}}$. Such a high velocity can be achieved if the pulsar experienced a relatively high natal kick. 
Finally, we discuss the binary companion properties that are most consistent with the observational properties of G292.0+1.8 and predictions for future binary companion searches.

\acknowledgments

We would like to thank Frank Winkler for useful comments on the manuscript and Vikram Dwarkadas for an insightful discussion about SNRs evolving in wind density profiles. We also thank the anonymous referee for constructive feedback that helped improve the manuscript.

\bibliographystyle{apj}

\end{document}

%% file: tab1.tex
\begin{deluxetable}{lccccc} 
\tablecolumns{6} \tablewidth{0pc} \tablecaption{\label{1Dparam}1D Model Parameters that produce the radius of the SNR, RS, and PWN}
\tablehead{
\colhead{Run} & \colhead{$\rm M_{ej}$} & \colhead{$\rm E_{51}$} & \colhead{$\rm n_{0}$} & \colhead{CSM Mass} & \colhead{ZAMS Mass} \\
\colhead{} & \colhead{(\msun)} & \colhead{($10^{51}$~erg)} & \colhead{($\rm cm^{-3}$)} &  \colhead{(\msun)} &  \colhead{(\msun)}
}

\startdata

1 & 1.4 & 0.3 & 0.212  & 10.0 & 13.0  \\
2 & 2.0 & 0.4 & 0.30 & 14.1 & 17.7   \\
3 & 2.5 & 0.5 & 0.35 & 16.4 & 20.5  \\
4 & 3.0 & 0.65 & 0.40 & 18.8 & 23.4  \\
5 & 5.0 & 1.0 & 0.69 & 32.4 & 39.0 \\
\enddata
\tablecomments{1-D model runs for the evolution of a PWN inside an SNR that produce the observed radius of the SNR shell, the reverse shock, and the PWN. The age for all models is 2500~years. The listed parameters are the ejecta mass ($M_{ej}$), explosion energy ($E_{51}$),average ambient density ($n_0$), swept-up CSM mass (ambient density multiplied by the volume of the SNR), ZAMS mass (sum of $M_{ej}$, CSM mass, and the neutron star mass of 1.6~$\rm M_{\odot}$). It can be seen that the ZAMS masses are implausibly high compared to $M_{ej}$ for runs 3--5.}
\end{deluxetable}

%% file: tab2.tex
\begin{deluxetable}{llccc}
\tablecolumns{5} \tablewidth{20pc} \tablecaption{\label{model_param} MODEL PARAMETERS} 
\tablehead{\colhead{Parameter} & \colhead{Description \hspace{10mm} Run:} & \colhead{1} & \colhead{3} & \colhead{5}} 
\startdata
$D$ (kpc) & SNR distance & 6.0 & \nodata & \nodata \\
$R_{SNR}$ (pc) & SNR radius & 7.7 & \nodata & \nodata \\
$R_{RS}$ (pc) & RS radius & 3.8 & \nodata & \nodata \\
$R_{PWN}$ (pc) & PWN radius & 3.8 & \nodata & \nodata \\
$t$ (1000 yr)  & SNR age & 2.5  & \nodata & \nodata \\
$M_{ej}$ ($\rm M_{\odot}$) & SN ejecta mass & 1.4 & 2.5 & 5.0\\
$n_0$ ($\rm cm^{-3}$) & Ambient density & 0.212 & 0.35 & 0.69\\
$E_{51}$ ($\rm 10^{51}\:erg$) & Explosion energy & 0.3 & 0.5 & 1.0 \\


$v_p$ (km/s) & Pulsar velocity & 500  & \nodata \\
$\dot{E_0}$ ($10^{38}$erg/s) & Initial spin-down lum. & 6.8 & 6.8 & 6.2 \\
$\dot{E}$ ($10^{37}$erg/s) & Current spin-down lum. & 1.2 & \nodata & \nodata \\
$n$ & Pulsar braking index & 3.0 & \nodata & \nodata \\
$\tau_0$ (1000 yr) & Spin-down timescale & 0.38 & 0.38 & 0.40 \\
$\tau_c$ (1000 yr) & Characteristic age & 2.9 & \nodata & \nodata \\
$B$ ($\rm \mu G$) & PWN magnetic field & 13 & 16 & 16 \\
\enddata
\tablecomments{The ejecta mass $M_{ej}$, pulsar braking index, initial spin-down luminosity $\dot{E_0}$, the initial spin-down timescale $\tau_0$, and the SNR age were adjusted to produce the observed SNR, RS, and PWN radii for an assumed distance of 6.0~kpc.}
\end{deluxetable}

%% file: tab3.tex
\begin{deluxetable*}{lcccc}
\tablecolumns{5} \tablewidth{0pc} \tablecaption{\label{ir_lines}IRS LINE FITS}
\tablehead{
\colhead{Line ID} & \colhead{Line Center} & \colhead{Line Flux} & \colhead{FWHM} & \colhead{ Velocity Shift} \\
\colhead{} & \colhead{(\micron)} & \colhead{($10^{-6} erg/cm^{2}/s/sr$)}  & \colhead{(\micron)}  & \colhead{($km/s$)}
}
\startdata
\sidehead{Position 1}
$[$\ion{Ne}{2}$]$ (12.8135) & 12.8355 $\pm$ 0.0008 & 55.6 $\pm$ 0.1& 0.1318 $\pm$ 0.0002 & 515 $\pm$ 24 \\
$[$\ion{Ne}{5}$]$ (14.3217) & 14.371 $\pm$ 0.003 & 5.6 $\pm$ 0.3 & 0.31 $\pm$ 0.01 & 1030 $\pm$ 70 \\
$[$\ion{Ne}{3}$]$ (15.5551) & 15.5710$\pm$ 0.001 & 15.7 $\pm$ 0.3 & 0.157 $\pm$ 0.003 & 300 $\pm$ 20 \\
$[$\ion{Ne}{5}$]$ (24.3175) & 24.366 $\pm$ 0.016 & 1.2 $\pm$ 0.2 & 0.30 $\pm$ 0.04 & 600 $\pm$ 200 \\
$[$\ion{O}{4}$]$ (25.8903) & 25.906 $\pm$ 0.003 & 14.6 $\pm$ 0.3 & 0.361 $\pm$ 0.006 & 190 $\pm$ 30 \\
$[$\ion{S}{3}$]$ (33.4810) & 33.45 $\pm$ 0.02 & 2.0 $\pm$ 0.4 & 0.32 $\pm$ 0.04 & -300 $\pm$ 180 \\
$[$\ion{Si}{2}$]$ (34.8152) & 34.75 $\pm$ 0.01 & 5.6 $\pm$ 0.5 & 0.37 $\pm$ 0.02 & -550 $\pm$ 100 \\
\sidehead{Position 2}
$[$\ion{Ne}{2}$]$ (12.8135) & 12.8468 $\pm$ 0.0002 & 37.4 $\pm$ 0.2 & 0.1287 $\pm$ 0.0005 & 780 $\pm$ 10 \\
$[$\ion{Ne}{5}$]$ (14.3217) & 14.378 $\pm$ 0.004 & 1.0 $\pm$ 0.1 & 0.12 $\pm$ 0.01 & 1170 $\pm$ 80 \\
$[$\ion{Ne}{3}$]$ (15.5551) & 15.592 $\pm$ 0.001 & 13.7 $\pm$ 0.4 & 0.157 $\pm$ 0.003 & 700 $\pm$ 20 \\
$[$\ion{S}{3}$]$ (18.7130) & 18.732$\pm$ 0.01 & 1.6 $\pm$ 0.3 & 0.16 $\pm$ 0.02 & 300 $\pm$ 150 \\
$[$\ion{Ne}{5}$]$ (24.3175) & 24.39 $\pm$ 0.02 & 1.4 $\pm$ 0.2 & 0.34 $\pm$ 0.04 & 880 $\pm$ 20 \\
$[$\ion{O}{4}$]$ (25.8903) & 25.953 $\pm$ 0.002 & 14.0 $\pm$ 0.3 & 0.354 $\pm$ 0.005 & 730 $\pm$ 30 \\
$[$\ion{S}{3}$]$ (33.4810) & 33.40 $\pm$ 0.04 & 2.3 $\pm$ 0.6 & 0.5 $\pm$ 0.1 & -740 $\pm$ 330 \\
$[$\ion{Si}{2}$]$ (34.8152) & 34.820 $\pm$ 0.009 & 6.9 $\pm$ 0.6 & 0.36 $\pm$ 0.03 & 40 $\pm$ 80 \\


\enddata
\tablecomments{Emission lines measurements from \textit{Spitzer} IRS spectra extracted from two regions on the PWN (shown as yellow rectangles in Figure~\ref{ir_lines}). Listed uncertainties are 1-$\sigma$ statistical uncertainties from the fit and do not include the IRS wavelength and flux calibration uncertainties.}
\end{deluxetable*}

%% file: tab4.tex
\begin{deluxetable}{lcrrrrrr}
\tablecolumns{8} \tablewidth{20pc}\tablecaption{Observed and Predicted Normalized Line Ratios \label{shock_table}}
\tablehead{    & & \multicolumn{2}{c}{Observed} &\multicolumn{4}{c}{Models}  \\
    & & & & 1 & 1 & 1 & 10  \\
Ion & $\lambda$ & Pos 1 & Pos 2 & 100 & 125 & 150 & 125}
\startdata
Ne II &  12.8 $\mu m$ &   380  &    270  &      1 &  $<1$&    1  &          6   \\
Ne V  &  14.3 $\mu m$ &    38  &      7  &     16 &   33 &   48  &         81   \\
Ne III&  15.5 $\mu m$ &   107  &     98  &     44 &   33 &   51  &        145   \\
S III &  18.7 $\mu m$ & $<68$  &     11  &    $<1$&  $<1$&  $<1$ &        $<1$  \\
Ne V  &  24.3 $\mu m$ &    79  &     10  &     30 &   67 &   88  &        100   \\
O IV  &  25.9 $\mu m$ &   100  &    100  &    100 &  100 &  100  &        100   \\
S III &  33.5 $\mu m$ &    14  &     16  &    $<1$&  $<1$&  $<1$ &        $<1$  \\
Si II &  34.8 $\mu m$ &    38  &     49  &    $<1$&  $<1$&  $<1$ &        $<1$  \\
      &        &        &         &        &      &       &     \\
O II  &  3737 \AA &   303  &      -  &      6 &    5 &    6  &          3   \\
Ne III&  3869 \AA &    19  &      -  &     45 &   29 &   31  &         40   \\
O III &  4363 \AA &     8  &      -  &      5 &    2 &    3  &          2   \\
O III &  5007 \AA &   100  &      -  &    100 &  100 &  100  &        100   \\
O I   &  6300 \AA &     8  &      -  &    $<1$&  $<1$&  $<1$ &        $<1$  \\
S II  &  6725 \AA &    17  &      -  &    $<1$&  $<1$&  $<1$ &        $<1$  \\
O II  &  7325 \AA &     8  &      -  &    $<1$&  $<1$&  $<1$ &        $<1$  \\
I(O IV)$^a$ &   & 1.46e-9 & 1.40e-9 & \\
\enddata
\tablecomments{$^a~~\rm erg ~cm^{-2}~s^{-1}~sr^{-1}$}
\end{deluxetable}